\documentclass[reprint,amsmath,amssymb,aps,prd, floatfix]{revtex4-1}
\usepackage{graphicx} % Required for inserting images
\usepackage{subcaption}
\usepackage{caption}
\usepackage{siunitx} % For the 'S' column type
\usepackage{tabularx}
\usepackage{geometry}
\usepackage[table]{xcolor}
\usepackage{multirow}
\usepackage{booktabs}
\usepackage{colortbl} 

\usepackage[nolist]{acronym}

\newcommand{\msun}{\ensuremath{M_\odot}}
\newcommand{\mchirp}{\ensuremath{\mathcal{M}}}

\begin{document}

\title{Optimized Search for a Binary Black Hole Merger Population in LIGO-Virgo O3 Data}

\author{Praveen Kumar}
\author{Thomas Dent}
\affiliation{IGFAE, Campus Sur, Universidade de Santiago de Compostela, 15705 Santiago de Compostela, Spain}
\date{\today}

\begin{abstract}
Maximizing the number of detections in matched filter searches for \ac{CBC} \ac{GW} signals requires a model of the source population distribution.  In previous searches using the PyCBC framework, sensitivity to the population of \ac{BBH} mergers was improved by restricting the range of filter template mass ratios and use of a simple one-dimensional population model.  However, this approach does not make use of our full knowledge of the population and cannot be extended to a full parameter space search. 
Here, we introduce a new ranking method, based on \ac{KDE} with adaptive bandwidth, to accurately model the probability distributions of binary source parameters over a template bank, both for signals and for noise events. 
%It considers signal and noise rate dependence on masses and spins via the signal and template KDE. It represents a significant evolution from the previously used O3 \ac{BBH} statistic.
We demonstrate this ranking method by conducting a search over LIGO-Virgo O3 data for \ac{BBH} with unrestricted mass ratio, using a signal model derived from previous significant detected events.  We achieve over $10\%$ increase in sensitive volume for a simple power-law simulated signal population, compared to the previous \ac{BBH} search. 
Correspondingly, with the new ranking, 8 additional candidate events above an \ac{IFAR} threshold $0.5\,$yr are identified. 
\end{abstract}

\maketitle

\acrodef{GW}{gravitational wave}
\acrodef{CBC}{compact binary coalescence}
\acrodef{BBH}{binary black hole}
\acrodef{BNS}{binary neutron star}
\acrodef{NSBH}{neutron–star--black-hole}
\acrodef{IMBH}{intermediate mass black hole}
\acrodef{KDE}{kernel density estimation}
\acrodef{awKDE}{adaptive width KDE}
\acrodef{SNR}{signal to noise ratio}
\acrodef{FAR}{false alarm rate}
\acrodef{IFAR}{inverse false alarm rate}
\acrodef{PE}{parameter estimation}

\section{Introduction}
\label{sec:intro}
%\paragraph{Discuss catalogs of CBC detections, general search methods}
The catalogue of detected \ac{GW} signals from compact binary mergers since the start of the Advanced detector era~\cite{LIGOScientific:2014pky,VIRGO:2014yos,Aso:2013eba} reached little short of 100 at the end of the O3 observing run~\cite{LIGOScientific:2021djp}.  Given the relatively low \acp{SNR} of such signals, currently inevitable due to extremely small \ac{GW} amplitudes despite advanced technologies, such signals cannot always be unambiguously distinguished from detector noise fluctuations.  The presence of non-Gaussian noise artifacts (e.g~\cite{LIGO:2021ppb}) also complicates the reliable identification of signals~\cite{Davis:2022cmw}. 

In drawing astrophysical inferences from such catalogs, one may wish to limit the expected overall contamination by noise events in a sample, or to impose that any given event of interest is signal rather than noise with sufficiently high confidence.  The first of these objectives is addressed by the calculation of \ac{FAR}, based on a ranking of detected event properties (the \ac{SNR} and various measures of signal consistency) from least to most signal-like.  Search methods and algorithms, including the choice of this ranking, may be optimized to maximize the number of events detected with \ac{FAR} less than a given threshold.  As detailed in~\cite{Biswas:2012tv,cannon2015likelihood,davies2020extending}, an optimal ranking is given by (any monotonic function of) the ratio of the likelihood of the measured data under the signal hypothesis to its likelihood under the noise hypothesis.  In practice we do not know either of these likelihoods exactly, so various approximations are applied.  In this work, in the context of a matched-filter binary merger search, we will consider the dependence of these likelihoods on binary masses and spins, and the consequences for search optimization, in more detail.

%\paragraph{Introduce problem of optimal search: maximize detections at fixed (I)FAR}
In such a search, in contrast to the case of a fixed, known signal, we are faced with a composite hypothesis: 
a binary merger \ac{GW} signal has many parameters with \emph{a priori} unknown values.  These can be separated into intrinsic parameters -- binary component masses and spins -- and extrinsic parameters, describing the relative spacetime positions and orientations of the observer and source.  To optimize search sensitivity, the likelihood ratio for the signal hypothesis \emph{vs}.\ the noise hypothesis (which in the ideal case is a Gaussian function of the matched filter \ac{SNR}) must be marginalized over both sets of parameters, using suitable prior distributions.  These ``optimal'' parameter priors in fact describe an astrophysical source population distribution~\cite{Dent:2013cva}. 
We then face the issue that this source distribution is partly unknown: while the distribution over binary orientation, direction from Earth and merger time can be found simply from symmetry arguments, the distribution over component masses and spins is only constrained by previous observations, or by theoretical/modelling expectations. 

In matched-filter searches, the technical problem of unknown source parameters is addressed by searching over large number of templates (i.e.\ candidate signal waveforms), arranged in order to minimize the loss of \ac{SNR} for any possible signal within a given parameter range (see e.g.~\cite{canton2017designing,Roy:2017qgg,Hanna:2022zpk}). 
The range, i.e.\ boundaries, of such a template bank have first to be set, generally by considering both the prior range of possible signals, and technical constraints such as computational cost: for instance searches for binaries with component masses below 1\,\msun require very large numbers of templates~\cite{LIGOScientific:2018glc}. 
To optimize search sensitivity for a given population, we should then ``weight'' each template (adjusting its ranking at a given \ac{SNR}) using its relative rate of detections, i.e.\ by the density of detectable signals over the bank parameters, divided by the density of templates~\cite{Dent:2013cva}.  Broadly, our aim in this work is to provide an approximate estimate of the signal density via a \ac{KDE} of previously identified, significant merger events.  The combination of the signal \ac{KDE} with a \ac{KDE} of the search templates produced with the same method will yield the optimal weighting, as far as it can presently be determined. 

%\paragraph{Link to correct classification of signal/noise by ranking / FAR / p\_astro}
The relative ranking of events occurring in different templates serves two apparently distinct goals.  First, to maximize the search sensitivity, i.e.\ the expected number of detected signals at a fixed \ac{FAR} threshold.  Second, returning to the question of confidence in an individual event, we wish the ranking to indicate accurately the relative probabilities of signal (astrophysical) \emph{vs}.\ noise (terrestrial) origin: a given ranking statistic value should correspond to a constant odds of signal \emph{vs}.\ noise.  Both of these goals are ideally achieved if the ranking accounts for the true distribution of signals, but as this is unknown, our strategy is to substitute a best estimate based on previous observations. 

%\paragraph{History of ranking over mass/spin} 
In searches performed before or soon after the first \ac{GW} signal detection~\cite{LIGOScientific:2016aoc}, there was little or no direct, or even indirect indication of the true binary mass and spin distribution, beyond the expectation of some nonzero rate of \ac{BNS} and \ac{NSBH} mergers~\cite{LIGOScientific:2010nhs}.  Hence, either a ``uninformative'' or ``template-agnostic'' ranking was used, effectively treating each template on an equal basis, or templates were weighted only on the basis of their different responses to detector noise transients~\cite{LIGOScientific:2016vbw,LIGOScientific:2016dsl,Nitz:2017svb}.  

As pointed out in~\cite{Dent:2013cva}, such rankings without a prior weighting are implicitly optimized for a distribution of signals (above a given \ac{SNR} threshold) which mirrors that of templates.  Given the frequency dependence of the binary signal phase and of detector sensitivities, the distribution of templates is heavily skewed towards much higher densities at low values of chirp mass $\mchirp = (m_1m_2)^{3/5}(m_1 + m_2)^{-1/5}$, where $m_1$ and $m_2$ are the binary component masses.  The density of templates over component spins is also generally somewhat higher towards extreme (near unity) dimensionless spin magnitudes.  Both of these trends are very different from the population of detectable merging binaries that has emerged in more recent observations, which is rather centered on moderate or large component masses from $\sim\!10\,\msun$ through $\sim\!60\,\msun$ or more, with generally low (of the order of $0.1$) effective orbit-aligned spins~\cite{LIGOScientific:2020kqk,KAGRA:2021duu}. 

The ``template-agnostic'' ranking used in searches up to the second LIGO-Virgo observing run (O2)~\cite{LIGOScientific:2018mvr} thus did not optimize detection probability, either for the actual population of merger signals or an approximate population model. 
Various strategies for improvement were deployed to produce event catalogues from O3: a ranking incorporating a prior signal distribution uniform over $\log m_1$, $\log m_2$ was used in the GstLAL search~\cite{Fong:2018elx,LIGOScientific:2020ibl,LIGOScientific:2021usb,LIGOScientific:2021djp}, accounting for template density effects, thus maintaining broad search sensitivity, though not necessarily a realistic estimate of the astrophysical population.  PyCBC-based searches pursued a different strategy: both a ``template-agnostic'' search over a broad parameter space, and a separate ``BBH'' search over a restricted range of template component masses ($m > 5\,\msun$) and mass ratios ($q \equiv m_2/m_1 > 1/3$) were run, with the BBH search using a ranking that approximates the relative densities of signals and templates as a power of \mchirp~\cite{LIGOScientific:2020ibl,LIGOScientific:2021djp}.  While this hybrid search strategy -- comparable to approaches used for independent searches of open data~\cite{Nitz:2018imz,Nitz:2020oeq,Nitz:2021uxj,Venumadhav:2019tad,Venumadhav:2019lyq} -- yields generally high sensitivity, it is unduly complicated: such setups imply abrupt (step-function) changes of the signal and template densities over binary mass parameters, and do not readily allow for more realistic modelling of the signal distribution.  

Searches using ``template-agnostic'' rankings also generally result in biased estimates of probability of astrophysical origin $p_\mathrm{astro}$.  An alternative approach pursued by~\cite{Andres:2021vew} for O3 catalogs~\cite{LIGOScientific:2021usb,LIGOScientific:2021djp} 
applied a broad population model to the estimate of $p_\mathrm{astro}$, but without using this model in the search ranking.  For the PyCBC search, such bias was partly addressed by estimating signal and noise event densities within separate template bins~\cite{LIGOScientific:2021usb,
LIGOScientific:2021djp}.  This binning approach, though, implies piecewise-constant estimates of signal density, which we do not expect to be a good fit to the actual distribution; moreover a binned estimate %of the signal distribution 
over more than one dimension is likely to be dominated by counting uncertainties, given the relatively small number of confidently detected events. 

Here, we pursue more sophisticated methods, which will ultimately allow a unified approach to candidate ranking within a single search.  We aim to retain the advantages in sensitivity and accuracy of $p_\mathrm{astro}$ estimation of a statistic that incorporates a realistic population prior over stellar-mass merging binaries, without imposing arbitrary or hard boundaries within a broad search space.  Ranking statistics incorporating prior population models have already been applied in a range of other scientific contexts, including a focused search for \ac{BNS} coalescences~\cite{Magee:2019vmb}, search for binaries with sub-solar-mass components~\cite{phukon2021hunt}, and searching for gravitationally lensed counterpart signals of known \ac{BBH}~\cite{li2023tesla}. 

This paper is structured as follows: Section \ref{sec:KDE_calculation} explains how we calculate KDEs in a three-dimensional space for detected signals and template banks, and how the resulting density estimates are implemented in the PyCBC ranking statistic.  
Section~\ref{sec:search_of_ligo} describes our search for BBH using this ranking in LIGO-Virgo O3 data: in Sec~\ref{sec:search_configuration} we explain the analysis configuration, in Sec.~\ref{sec:search_sensitivity} we compare its sensitivity to previous PyCBC searches, and in Sec.~\ref{sec:search_results} we give complete search results from our analysis and compare these with previous candidate lists. 
Section \ref{sec:discussion} summarizes the implications of our results and discusses further technical issues and possible developments.

\section{\ac{KDE}-based ranking Statistic for the PyCBC search}
\label{sec:KDE_calculation}

\subsection{Technical context}
The PyCBC offline search pipeline, built using the PyCBC software package, \cite{Allen:2005fk,DalCanton:2014hxh,usman2016pycbc, Nitz:2017svb},
plays an important role in identifying candidate \ac{GW}s from \ac{CBC}s. It uses a matched filtering technique to correlate observed data with template waveforms to calculate the \ac{SNR}. Triggers are generated in each detector separately by applying a threshold and clustering to the resulting \ac{SNR} time series. To improve the distinction between \ac{CBC} signals and non-Gaussian noise transients, a $\chi^{2}$ test is performed for each trigger \cite{Allen:2004gu}. The matched-filter \ac{SNR} is re-weighted by the value of the $\chi^{2}$ statistic; for signals matching our templates, the reduced $\chi^{2}$ is expected to be near unity.
 
The search checks for any triggers during periods of  instrumental or
environmental artifacts and veto them. For a candidate event to be considered, it must have a consistent arrival time in different detectors, within a \ac{GW} travel time window. The triggers also need to match the same template in each detector, ensuring consistency. To minimize false alarms, a coincidence test is conducted across all detectors, verifying parameter consistency \cite{usman2016pycbc}. Triggers passing both time and parameter coincidence tests are labeled as candidate events. 

The next step of the pipeline is to measure the \ac{FAR} as a function of detection statistic and use this to assign a statistical significance to candidate events. It is measured empirically by artificially shifting the time of triggers in one detector compared to another and then identifying the resulting (non-physical) coincident events. 
Repeating this analysis for many different choices of time shift creates a big set of background data that is employed to estimate the \ac{FAR} of the search. Since different templates in the bank can respond differently to detector noise, the search background isn't the same across all templates. 
The measured noise background is thus fitted as a function of the template parameters, to account for the variations of the noise distribution across the target signal space \cite{Nitz:2017svb}.

The ranking statistic is a fundamental aspect of the search pipeline, offering a systematic approach to evaluating the significance of candidate coincidences. It is a function that incorporates an array of factors such as matched filter \ac{SNR}, $\chi^{2}$ values from signal-based vetoes, as well as intrinsic and extrinsic parameters characterizing potential \ac{GW} signals \cite{Nitz:2017svb}. It serves as a crucial tool for distinguishing astrophysical \ac{GW} events from background noise. The calculation of significance also requires that candidate events are statistically independent. However, both noise and signals can create multiple triggers across the bank, leading to several related events in a short time. To ensure independence, the pipeline performs a final step of clustering: if more than one event happens within a fixed time window, only the one with the highest detection value is chosen as a candidate.  The same clustering step is also applied to events in the time-shifted analyses.  
To maximize our ability to distinguish GW signals in the detector data, it is crucial to have a ranking statistic that reflects our knowledge of the signal distribution. As discussed in Section \ref{sec:intro}, this ensures a reliable indication of the relative probabilities between signals and noise.

Here, we introduce a new method to assign ranking statistic based on \ac{KDE} with adaptive bandwidth selection. This method allows for more accurate estimation of the rate densities of signal and noise triggers, which are then used to calculate the optimal or likelihood ratio statistic as follows: 
\begin{equation}
    \Lambda_{SN}(\vec{k}) = \frac{r_{S}(\vec{k})}{r_{N}(\vec{k})},
\end{equation} 
where $r_{S}(\vec{k})$ and $r_{N}(\vec{k})$ are the rate densities of signal and noise triggers over $\vec{k}$ respectively, and the vector $\vec{k}$ includes an event's intrinsic and extrinsic parameters: 
\begin{equation}
    \vec{k} = \{ [\rho_{a}, \chi^{2}_{a}, \sigma_{a}], \vec{\theta}, [\mathfrak{A}_{ab}, \delta t_{ab}, \delta \phi_{ab}] \}. 
\end{equation}
Here $\rho_{a}, \chi^{2}_{a}$ and $\sigma_{a}$ are the trigger \ac{SNR}, signal-glitch discriminators, and template sensitivity respectively for each participating detector, denoted by $a$. Template intrinsic parameters, $\vec{\theta}$, comprise the binary component masses and spins. 
The network consistency parameters
$\mathfrak{A}_{ab}, \delta t_{ab}$, and $\delta \phi_{ab}$ are the amplitude ratio, time difference, and phase difference respectively 
between each detector pair labeled by $a \neq b$.  We write the signal rate density as $r_{S} (\vec{k}) \equiv \mu_{S} \hat{r}_{S}(\vec{k})$, where $\mu_{S}$ is an astrophysical coalescence rate per volume per time, assumed to be constant. 
To account for the vast dynamic range of expected rate densities, the logarithm of the ratio of signal and noise rate densities has been used \cite{davies2020extending}; 
%\begin{equation}
%    \mathfrak{R} = log \Lambda_{SN} + constant
%\end{equation}
%The equation gives the relation between ranking statistic, $\mathfrak{R}$ and optimal detection statistic, $\Lambda_{SN}(\vec{k})$. 
%\td{this sentence doesn't tell the reader anything beyond looking at the equation: if the constant is just derived from $mu_S$ I think it is not worth writing this as a displayed equation.}
thus, we obtain a ranking statistic
\begin{equation}
    \mathfrak{R} = \log \hat{r}_{S}(\vec{k}) - \log r_{N}(\vec{k}),
\end{equation}
which is equal to $\log \Lambda_{SN}$ up to a constant. 
The dependences of $\mathfrak{R}$ on $\rho_{a}, \chi^{2}_{a},\sigma_{a}$ and on the network consistency parameters $\mathfrak{A}_{ab}, \delta t_{ab}, \delta \phi_{ab}$ were already estimated in the previously derived ranking of~\cite{davies2020extending}, which we will denote as $\mathfrak{R}_0$. 

Here, we are interested in the dependence of the statistic on the template intrinsic parameters, $\theta$.  The rate of noise triggers \emph{per template}, which depends on both the properties of the template waveform and the data, is included as a term in $\mathfrak{R}_0$, but no explicit dependence on $\vec{\theta}$ was included.  As discussed in the Introduction, this choice implies that every template is equally likely to detect a signal at any given \ac{SNR} \cite{Dent:2013cva}.  With a more general choice of ranking, including an explicit model of the distribution of signals, we have
\begin{equation} \label{eq:R_and_R0}
    \mathfrak{R} = \mathfrak{R}_0(\vec{k}) + \log \frac{d_S (\vec{\theta})}{d_T (\vec{\theta})},
\end{equation} 
where the second term is the (log of) the ratio of signal and template densities over the space of masses and spins $\vec{\theta}$.

For producing catalog results on O3 data, the PyCBC search was run with two different configurations: Broad (covering the full parameter space) and \ac{BBH} (focused) \cite{LIGOScientific:2020ibl}. 
The broad search is designed to detect a diverse range of signals, exploring a wide spectrum of masses and spins. On the other hand, the \ac{BBH} search is focused specifically on the regions with mass ratios and component masses ranging over $1/3 \leq q \leq 1$, and $5 \msun \leq m_{1} \leq 350 \msun$ ($m_{2} \geq 5 \msun$), respectively. 
These two searches had various differences in configuration, most notably in the choices of ranking statistic.  In the broad search using $\mathfrak{R}_0$ as statistic, $d_S (\vec{\theta}) / d_T (\vec{\theta})$ is effectively set to $1$; however in the \ac{BBH} search, a nontrivial weighting is used. 
% the likelihood is adjusted by multiplying the signal rate. 
% TD  This is still not quite right.  As you can see from the algebra, the factor arises from the template distribution so I don't think 'multiplying the signal rate' is right 
The distribution of templates over chirp mass has an approximate dependence $\propto \mchirp^{-11/3}$, thus the weighting takes the inverse of this factor, giving a ranking statistic~\cite{Nitz:2020oeq}
%for O3 \ac{BBH}
\begin{equation}
    \mathfrak{R}_{\mathrm{BBH}} = \mathfrak{R}_0 + \frac{11}{3} \log \left(\frac{\mchirp}{\mchirp_*}\right),
    \label{BBH_stat_eqn}
\end{equation}
where $\mchirp_*$ is a constant reference mass.  Hence, 
higher mass templates receive more weight to ensure that quiet signals at high masses are not overshadowed by more numerous noise events at lower masses.  As a result, this statistic was approximately optimized for a signal distribution \emph{uniform} over $\mchirp$ (at a constant \ac{SNR}); compare the ranking used in~\cite{Magee:2019vmb}, optimized for a \ac{BNS} population following a Gaussian distribution over $\mchirp$. 

This O3 configuration, with two separate searches to cover the mass space, was effective at maintaining relatively high sensitivity, but has the disadvantage of complexity; there is also a difficulty in assessing the significance of any candidate, as the two searches both produce noise events, which may be partly correlated or in common with each other.  Here, we would like to both simplify the configuration by running one single search, and also maintain the advantage in sensitivity that the O3 \ac{BBH} search derived from its restricted range of $q$ and its 1-d estimate of the template distribution. 

To demonstrate our new approach, for computational cost reasons, we will conduct a search over only the stellar-mass \ac{BBH} space with component masses $m_1, m_2 \geq 5 \msun$ but \emph{without} any restriction on mass ratio; the corresponding ``unrestricted BBH bank'' is detailed later in Sec.~\ref{sec:search_configuration}.

\subsection{KDE evaluation and use in ranking statistic}
 
To develop a ranking statistic across a bank covering a wider range of mass ratios, we explore the dependence of signal/template densities over mass ratio and spin, which was neglected in the O3 \ac{BBH} ranking.  We now estimate signal and template densities over a three-dimensional space of $\log(\mchirp)$, $\eta$, and $\chi_\mathrm{eff}$, 
where $\eta$ represents the symmetric mass ratio $m_1 m_2 / (m_1 + m_2)^{2}$ and $\chi_\mathrm{eff}$ is the effective spin defined as $(s_{1z} m_1 + s_{2z} m_2) / (m_1 + m_2)$.  
Instead of the function of \mchirp\ used in the \ac{BBH} statistic, the term $d_S (\vec{\theta}) / d_T (\vec{\theta})$ is given as the ratio of \acp{KDE} calculated from signal and template points.  
The KDE ranking is thus given as
\begin{equation}
    \mathfrak{R}_{\mathrm{KDE}} = \mathfrak{R}_0 + \log d_S (\log \mchirp, \eta, \chi_\mathrm{eff}) - \log  d_T  (\log \mchirp, \eta, \chi_\mathrm{eff}). 
    \label{KDE_stat_eqn}
\end{equation}
The 3-dimensional \acp{KDE} thus incorporate a more accurate model of the signal and template / noise distributions over masses and spins than for O3 \ac{BBH}, with signal and template distributions modelled over chirp mass only. 
This not only allows us to calculate a ranking statistic more accurately, helping us better identify and understand astrophysical sources, but also serves our primary purpose of maximizing sensitivity or increasing the number of detections.

To evaluate $d_{S,T}(\log \mchirp, \eta, \chi_\mathrm{eff}$), we use \ac{awKDE} in preference over the fixed global bandwidth KDE, because the latter tends to over- or underestimate the width of features in the distribution, which may neglect structures of interest or introduce unphysical gaps in the estimated distribution.  Fixed bandwidth \ac{KDE} struggles to both accurately reconstruct small-scale features in densely populated regions, and supply sufficient smoothing to avoid artifacts in sparsely populated regions \cite{sadiq2022flexible}. In contrast, \ac{awKDE} addresses these limitations by dynamically adjusting the bandwidth according to local point density, allowing for a more accurate estimation.  %of the underlying distribution. 
A similar \ac{awKDE} technique for the one- and two-dimensional cases was demonstrated by \cite{sadiq2022flexible} and we extend the same approach to three dimensions. The algorithm calculates an \ac{awKDE} estimator $\hat{f}$, using as input the measured parameters $\vec{X}_{i}$ of a set of independent events labelled by $i$, 
via~\cite{scott2015multivariate} 
\begin{equation}
    \hat{f}(\vec{x}) = \frac{1}{n} \sum_{i=1}^{n} \frac{1}{h\lambda_i}K \left( \frac{\vec{x} - \vec{X}_{i}}{h\lambda_i} \right),
    \label{KDE_formula}
\end{equation}
where $K$ is the standard (in general multivariate) Gaussian kernel, 
\begin{equation}
    K (\Vec{y}) = \frac{1}{\sqrt{2\pi}} \exp \left( - \frac{1}{2} |\vec{y}|^2 \right),
\end{equation}
$n$ is the total number of samples and the term $h\lambda_{i}$ accounts for a local bandwidth. 
For the adaptive estimator, we aim to take into account the local density of events when choosing the kernel bandwidth: regions with higher (lower) densities will have narrower (broader) kernels applied. This helps to avoid excess variance artefacts in regions with low event density, by applying more smoothing there, % TD - I pointed out that in low density regions the problem is that the fixed BW has artefacts and too much variance, so we need more smoothing. 
and conversely provides higher precision in regions with high event density. 
To implement the adaptive KDE, a pilot density $\hat{f}_0$ is first calculated by setting $\lambda_i = 1$ for all $i$, i.e.\ a standard fixed bandwidth \ac{KDE}.  Then based on pilot density, the local bandwidth parameter is defined as 
 \begin{equation}
     \lambda_i = \left( \frac{\hat{f}_0(X_i)}{g} \right)^{-\alpha},  \qquad \log g = \frac{1}{n} \sum_{i=1}^{n} \log \hat{f}_0 (X_{i}),
\end{equation}
where $\alpha$ is the local bandwidth sensitivity parameter, with a value $0 < \alpha \leq 1$, and $g$ is a normalization factor. 
The method requires the %values of the 
initial global bandwidth $h$ and sensitivity parameter $\alpha$ values to be set: we assigned these using a 2-d grid search, with the figure of merit to be maximized being the total likelihood of the training samples evaluated via cross-validation~\cite{sadiq2022flexible}. 

\subsection{Validation of the KDE statistic}

%- We demonstrate the effect of the statistic, as a proof of principle, for a coarse BBH template bank covering an unrestricted range of mass ratios, and for a set of signals obtained only up to the first half of the O3 run (O3a). 

%- Describe how the sbbh2 bank was made (ie 3-OGC plus cut on m2)

%- Describe how we got the set of signal events and their parameters from 3-OGC (similar to the text for 4-OGC below). 

%- Say that we analyzed 7 out of the 9 time periods used for the PyCBC BBH search of O3 .. maybe state total calendar time.  Say that the search configuration is similar to the one used for the 3-OGC broad BBH search.  Say that we obtained results for both the mchirp-based O3BBH statistic and for the KDE statistic and we make a sensitivity comparison based on the BBH injection set supplied by LVK (discussed further in \ref{sec:search_sensitivity})~\cite{LIGOScientific:2021djp,ligo_scientific_collaboration_and_virgo_2023_7890437}. 

We demonstrate the effect of the statistic as a proof of principle for a coarse \ac{BBH} template bank covering an unrestricted range of mass ratios.  We used a template bank from the 3-OGC search~\cite{Nitz:2021uxj}: that search was carried out with a bank constructed in four parts using a stochastic placement method~\cite{harry2009stochastic}.
Out of these, we use the ``All BBH'' bank, which is laid out with a minimal-match value of 0.97 (see Fig.~2 of~\cite{Nitz:2021uxj}); however, we removed all templates with component masses $m_1, m_2 > 5\,\msun$, as the great majority of \ac{BBH} events have higher masses, and the computational cost of the study is greatly reduced by this cut. 
The signal events used are from the 3-OGC analysis \cite{Nitz:2021uxj}, covering up to the first half of the O3 run; as for the template bank, we impose the condition $m_1, m_2 > 5\,\msun$ on signal component masses, and additionally apply a threshold of $\mathrm{IFAR} > 0.5$ yr to select only significant signal events, as also in Sec.~\ref{sec:confi_optized_search}. 
We thus construct KDEs using a total of 11,207 templates, or 39 signal events respectively. 

\begin{figure*}[tbh]
    \centering
    \begin{minipage}{\textwidth}
        \centering
        \includegraphics[scale=0.11]{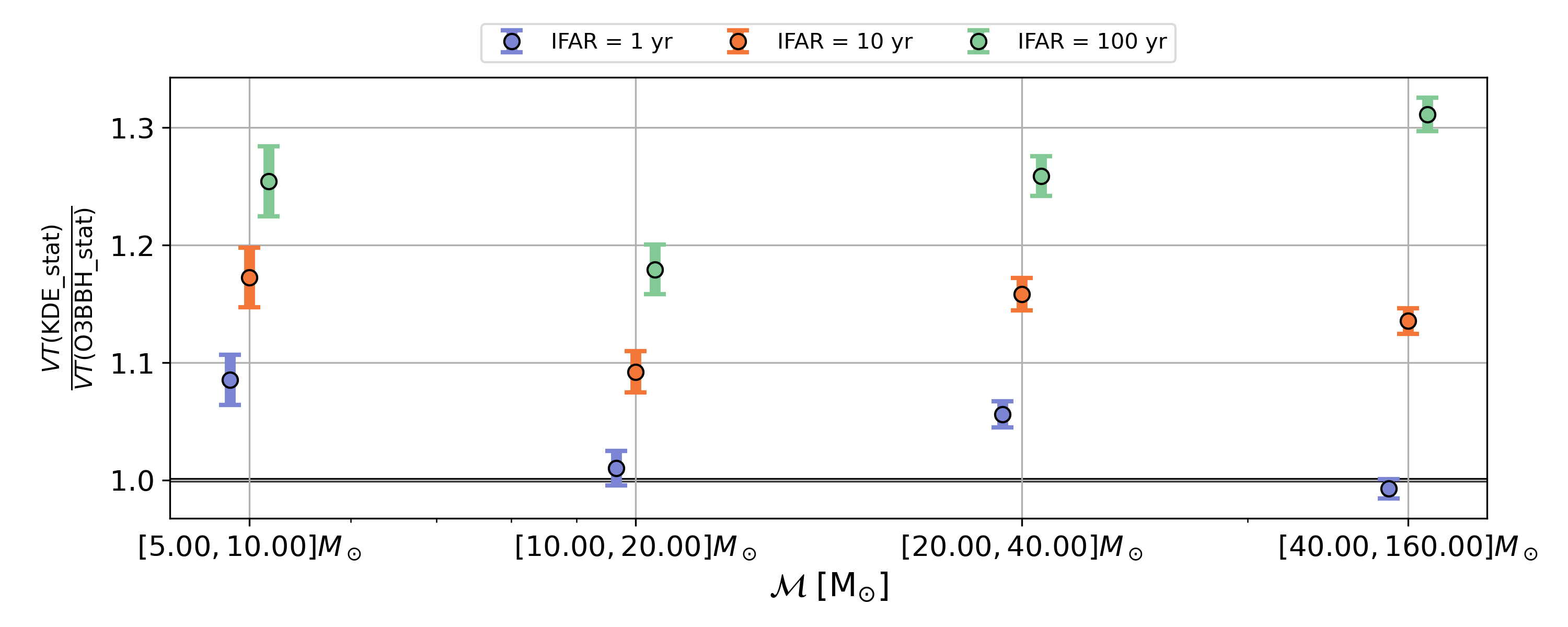}
    \end{minipage}
    \begin{minipage}{\textwidth}
        \centering
        \includegraphics[scale=0.106]{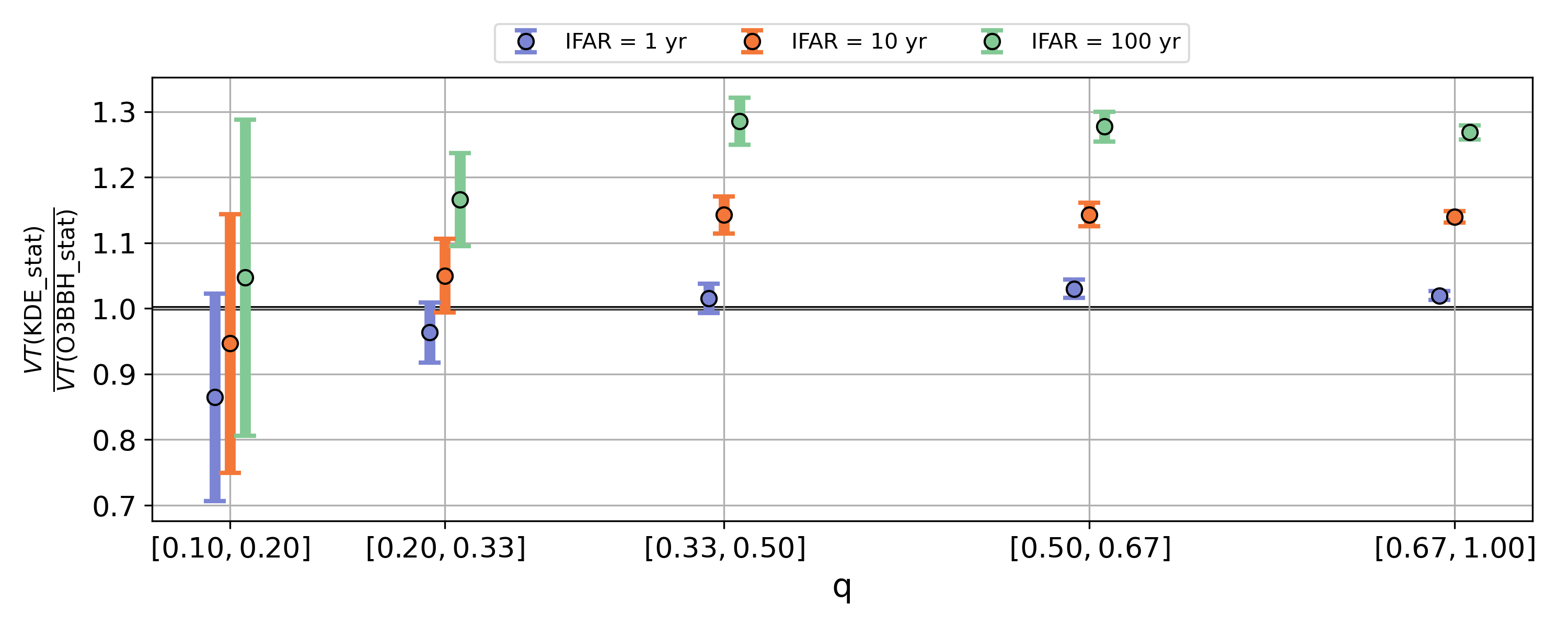}
    \end{minipage}
    \caption{Direct comparison of search sensitivity between two ranking statistics: \ac{KDE} and O3 \ac{BBH}, for the same bank and the same set of injections. %\td{this is not really two searches, it is the same search pipeline with two different statistics.}
    The top and bottom subplots show $\langle VT \rangle$ comparisons for injections divided into chirp mass bins and mass ratio bins respectively, at different \ac{IFAR} thresholds.}
    \label{vt_sbbh2}
\end{figure*}
We analyzed O3 data from 2019-04-01 to 2020-01-13 in order to compare results with the O3\ac{BBH} and \ac{KDE} ranking statistics.  
Detailed search configuration settings are discussed in Section \ref{sec:search_configuration}, and the suite of simulated signals (injections) analyzed to estimate sensitivity is described in detail in Section~\ref{sec:search_sensitivity}. 
%We ran the PyCBC search with both ranking statistics: O3\ac{BBH} and \ac{KDE}. After the search, 
A comparison of search sensitivity over bins in chirp mass and mass ratio, at various \ac{IFAR} thresholds, is shown in Fig.~\ref{vt_sbbh2}.  We observed higher sensitivity with the KDE statistic both when binned over $\mchirp$ and $q$, particularly at higher significance thresholds, except at lower $q$ values where the KDE statistic gives slightly lower sensitivity, within the uncertainty due to a finite set of injections.  This trend is expected, as the KDE statistic is designed to down-rank parameter ranges which are less represented (or absent) in the signal training set.

\subsection{Configuration for optimized O3 search}
\label{sec:confi_optized_search}

We now discuss specific choices in our calculation of signal and template \acp{KDE} used for the optimized search presented in the next section. 
For the signal case, we use events from the 4-OGC analysis of data up to the end of O3~\cite{nitz20234}, deriving our training points from the median \ac{PE} source masses and applying a factor $(1+z)$ based on the \ac{PE} median redshifts.  
For the template case, we use a stochastic \ac{BBH} bank with unrestricted mass ratio; all the templates have $m_2 > 5\,\msun$ and $m_1 > m_2$ (again corresponding to detector frame, i.e.\ redshifted, masses), extending up to masses of 500\,\msun; this bank is described in detail in Sec.~\ref{sec:search_configuration}.  
%and with unrestricted mass ratio.  
In the signal case, an additional condition is applied, requiring $\mathrm{IFAR} > 0.5$\,yr: we then impose the same minimum component mass of $5\,\msun$, resulting in a total of 59 events.  These conditions ensure that the events used in the analysis are \ac{BBH}-like.  
We then evaluated the \ac{KDE} using (\ref{KDE_formula}) over the three-dimensional space of $\log(\mchirp)$, $\eta$, and $\chi_\mathrm{eff}$ parameters of the template bank for both signal and template cases. 

\begin{figure*}[tbh]
    \centering
    \begin{minipage}[b]{0.32\textwidth}
        \includegraphics[trim=10 10 20 10, clip,width=\linewidth]{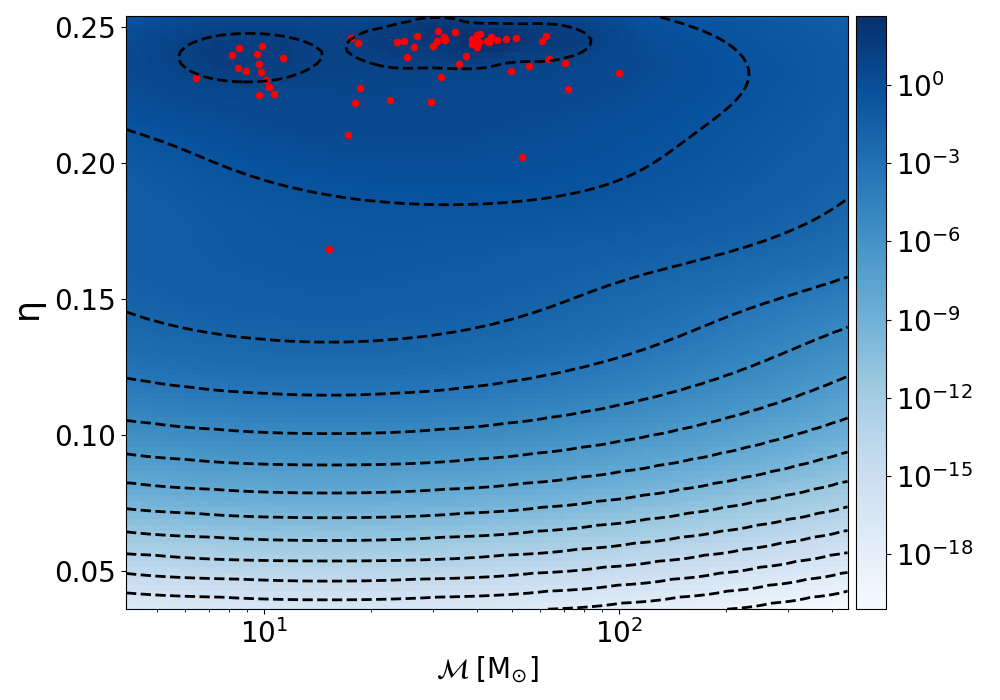}
    \end{minipage}
    \begin{minipage}[b]{0.32\textwidth}
        \includegraphics[trim=10 10 20 10, clip,width=\linewidth]{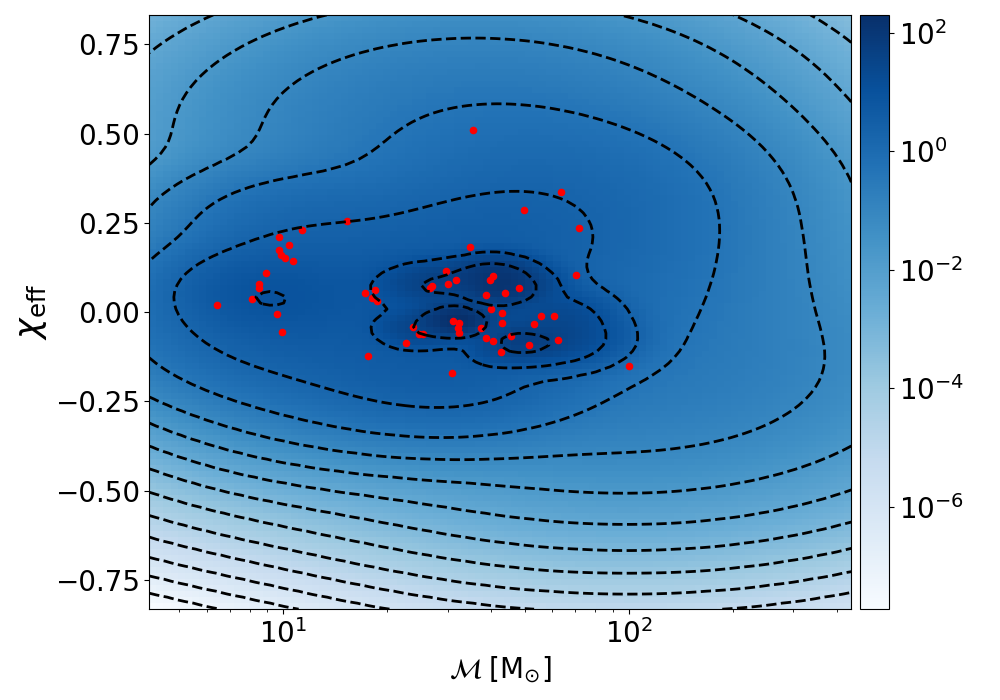}
    \end{minipage}
    \begin{minipage}[b]{0.32\textwidth}
        \includegraphics[trim=10 10 18 10, clip,width=\linewidth]{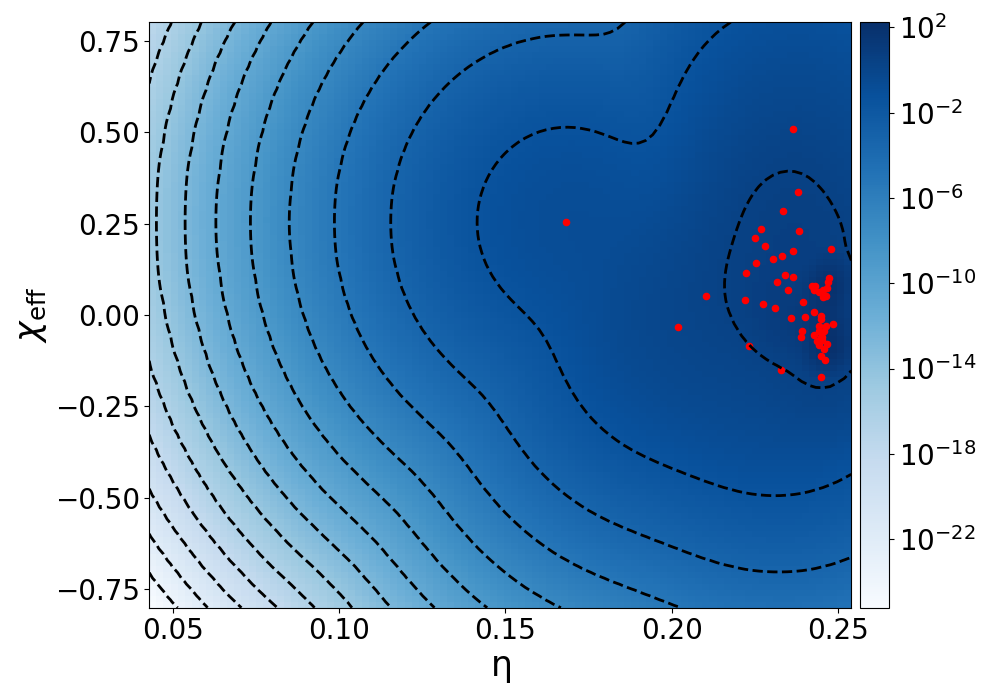}
    \end{minipage}

    \vspace{0.1cm}
    \begin{minipage}[b]{0.32\textwidth}
        \includegraphics[trim=10 10 20 10, clip,width=\linewidth]{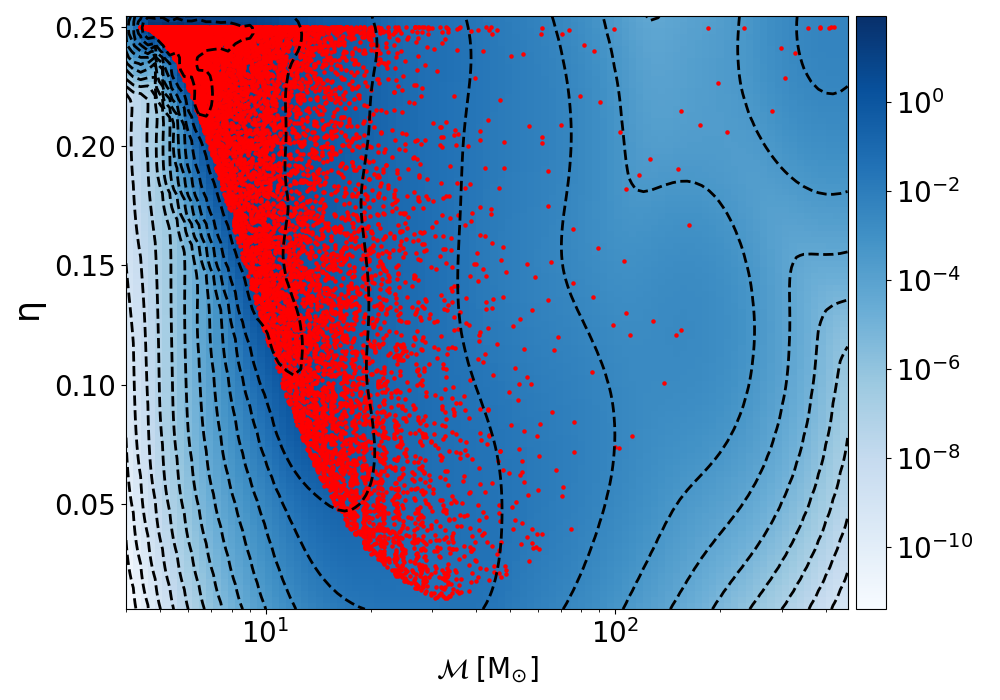}
    \end{minipage}
    \begin{minipage}[b]{0.32\textwidth}
        \includegraphics[trim=10 10 20 10, clip,width=\linewidth]{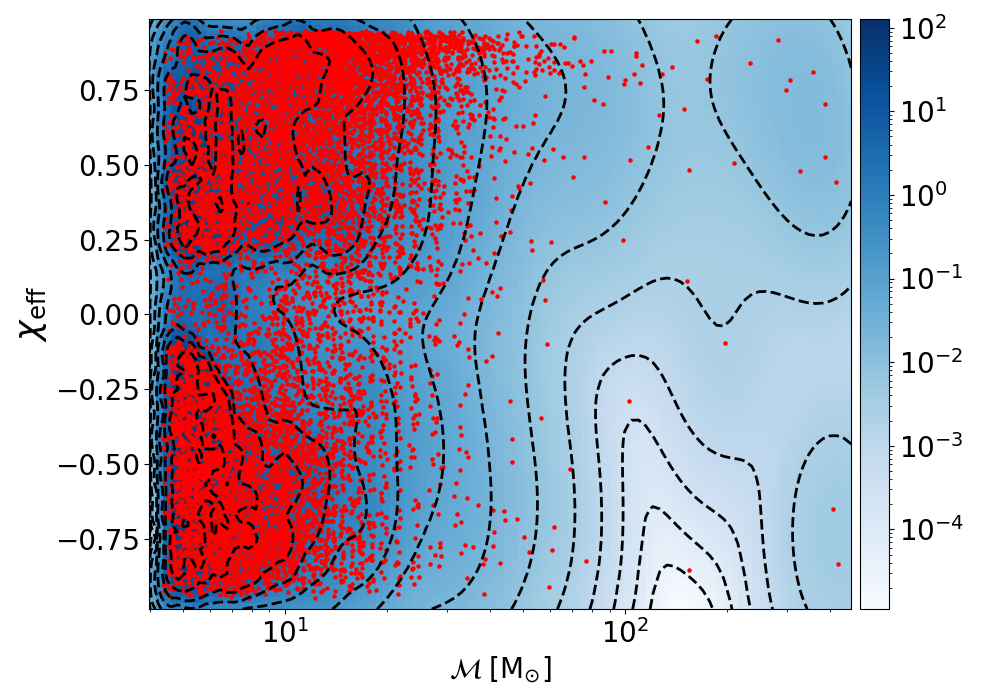}
    \end{minipage}
    \begin{minipage}[b]{0.32\textwidth}
        \includegraphics[trim=10 10 20 10, clip,width=\linewidth]{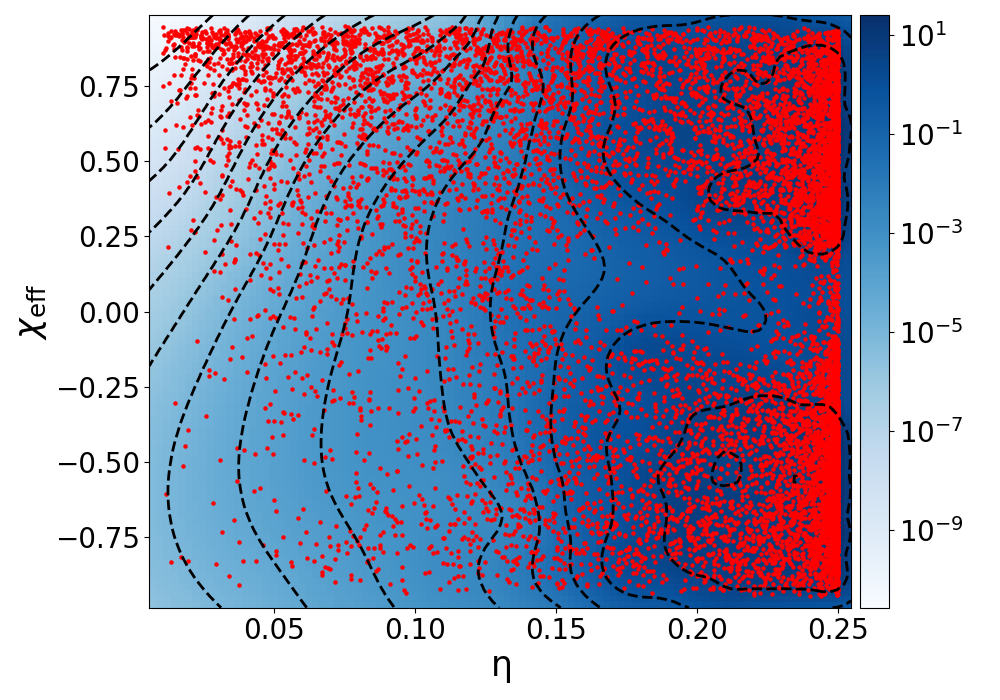}
    \end{minipage}
    \caption{Top / bottom row: The signal, resp.\ template \ac{KDE}, plotted over various 2-d ``slices'' through our 3-d parameter space. The red dots represent signals and templates in the signal and template KDE, respectively, used for training. 
    The values of the third parameter for these slice plots are $\chi_\mathrm{eff} = 0$, $\eta = 0.25$, and $\log(\mchirp) = 2$ in both rows, respectively.}
    \label{densities}
\end{figure*}

\begin{figure*}[tbp]
\centering
\begin{minipage}{0.307\textwidth} 
  \centering
  \includegraphics[trim=5 5 115 0, clip,width=\linewidth]{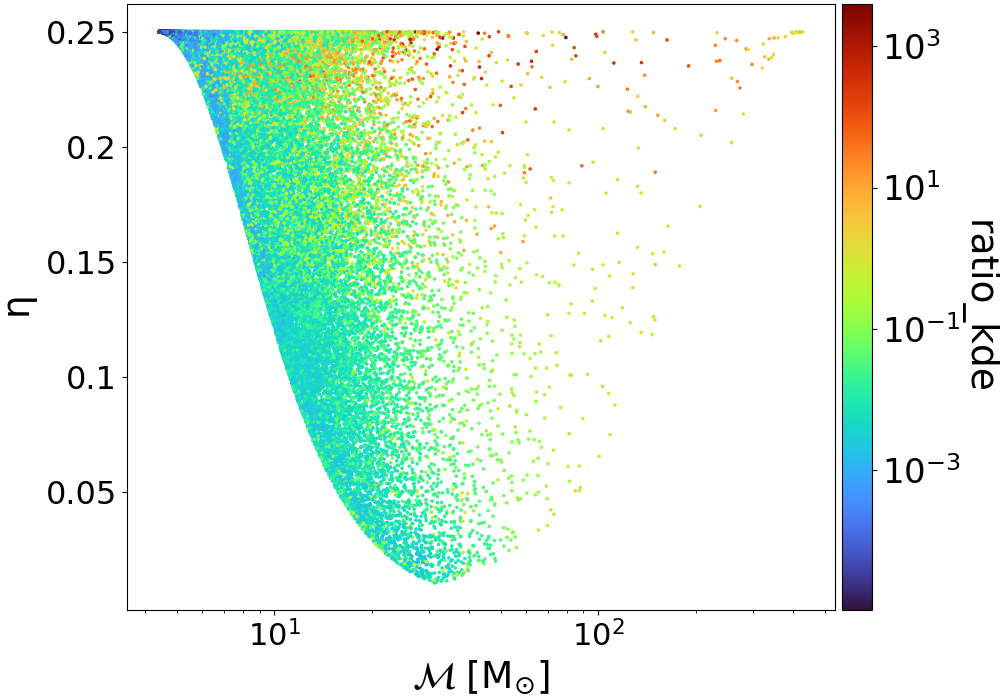}
\end{minipage}
\begin{minipage}{0.307\textwidth}
  \centering
  \includegraphics[trim=5 5 115 0, clip,width=\linewidth]{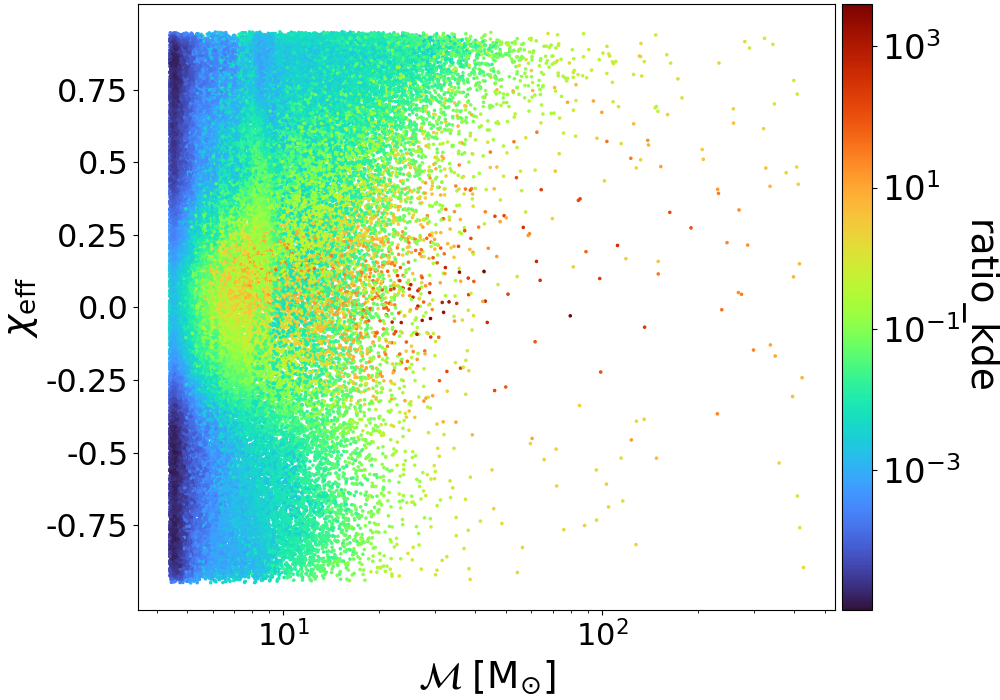}
\end{minipage}
\begin{minipage}{0.365\textwidth}
  \centering
  \includegraphics[trim=0 0 0 0, clip,width=\linewidth]{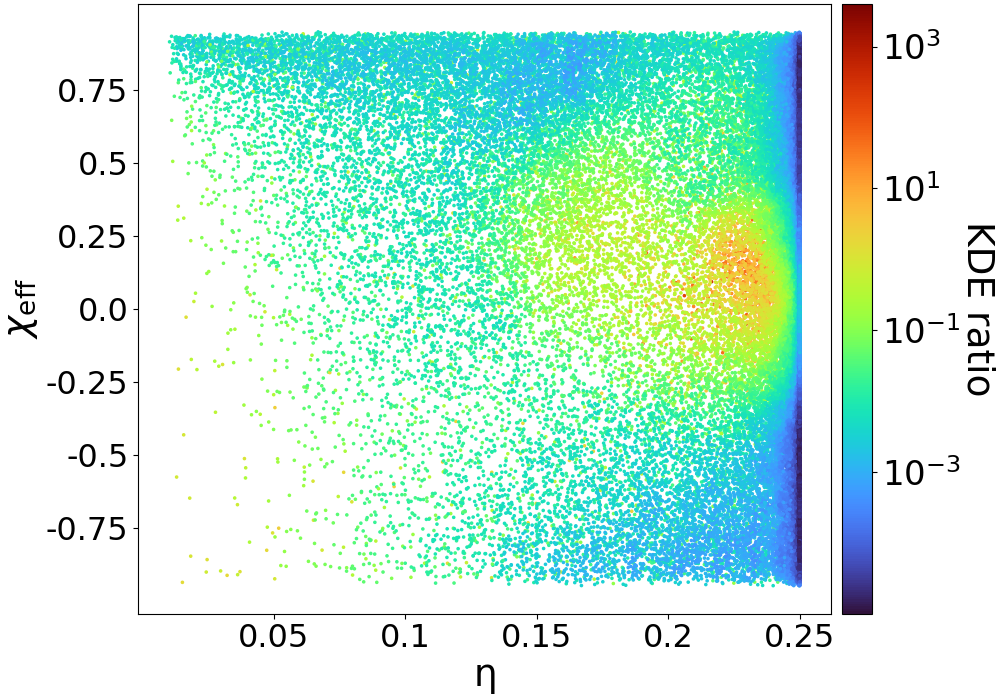}
\end{minipage}
\caption{Scatter plots between the different parameters used in the \ac{KDE}. The colorbar is a ratio of signal to template \ac{KDE}, evaluated at the template points.}
\label{ratio}
\end{figure*}

We show the resulting \acp{KDE} in Fig.~\ref{densities} for both the signal (top) and template (bottom) cases across different 2-d parameter space slices. 
% KDE already means 'estimate' or 'estimation' so we should not say 'estimates of KDE'
%across different space parameters are shown, covering a complete range of parameters in both scenarios. 
In the signal \ac{KDE} (top), there is a higher density of events 
concentrated around $\eta \approx 0.25$, $\chi_\mathrm{eff} \approx 0$, and $\mchirp \approx 40\,\msun$, while for the template \ac{KDE} (bottom) we observe various different trends; for instance the density of templates is higher towards nonzero $\chi_\mathrm{eff}$ values (both positive and negative).
Scatter plots of the ratio of signal to template \acp{KDE} 
at the template points are shown in Fig.~\ref{ratio}: 
These reflect that the highest density of detected events is at high values of $\mchirp$ and $\eta$ and near $\chi_\mathrm{eff} = 0$.

\section{Search of Ligo-Virgo O3 data}
\label{sec:search_of_ligo}

\subsection{Search configuration and sensitivity}
\label{sec:search_configuration}

To maintain high sensitivity while conducting only one search for BBH-like systems, in contrast to the previous PyCBC search strategy in O3, we employed an unrestricted \ac{BBH} bank in this work, covering a broad range of component masses (5 to 500\,\msun) and orbit-aligned component spins (-$0.949$ to $0.949$). 
The O3 BBH search achieved high sensitivity with a small number of templates, and so a low noise background, but at the cost of leaving the small-$q$ region to a separate search with worse template coverage, and of having two (mostly) independent sets of noise events over the two searches. Our approach aims to address these challenges by covering the entire range of $q$ in a single search, while maintaining sensitivity to the near-equal-mass \ac{BBH} population. 

\begin{figure*}[tbp]
\centering
\begin{minipage}{0.49\textwidth} 
  %\centering
  %\hspace{-0.45cm}
  \includegraphics[width=0.85\linewidth]{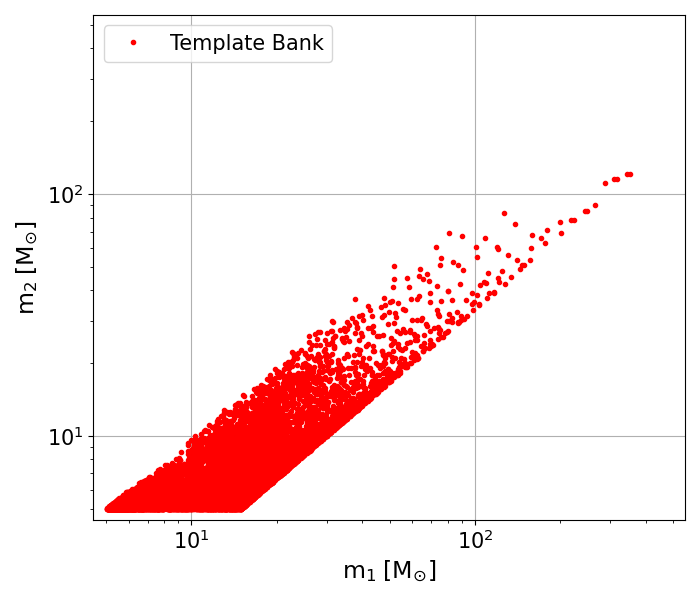}
\end{minipage}
\begin{minipage}{0.49\textwidth}
  %\centering
  %\hspace{-0.45cm}
  \includegraphics[width=0.85\linewidth]{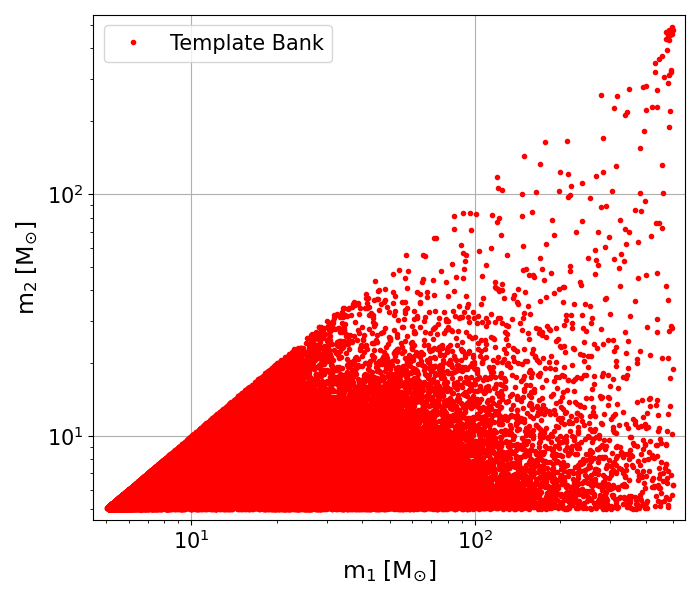}
\end{minipage}
\caption{Template placement for the O3 focused \ac{BBH} bank with restricted $q$ (17,094 templates, left) and our \ac{BBH} bank with unrestricted $q$ (64,184 templates, right).} 
\label{bank}
\end{figure*}

Our bank is generated using a brute force stochastic method with a minimal-match value of $0.985$ \cite{harry2009stochastic}, employing the $\mathrm{SEOBNRv4\_ROM}$ waveform approximant for templates, and using the same power spectral density estimate as in the 3-OGC search of O3a \cite{Nitz:2021uxj}. 
The chosen minimal match strikes a good balance between optimizing the computational speed and minimizing \ac{SNR} loss. 
While it might seem logical to increase the value to reduce \ac{SNR} loss risk, 
the major drawback of raising it is the increased computational cost: as the value goes up, the number of templates in the bank also increases. Thus, a balance is needed between the template bank's quality and the computational resources required. 
Fig.~\ref{bank} (right) shows our unrestricted BBH bank in the $m_1$-$m_2$ plane, for comparison with the O3 focused \ac{BBH} bank (left). 

We analyze the complete set of data from the O3 run, recorded and calibrated by the LVK collaborations, which is also accessible via the Gravitational Wave Open Science Center (GWOSC)\footnote{https://gwosc.org/GWTC-3/}: 
we used offline recalibrated (C01) strain data, applying available vetoes for invalid or substandard data quality~\cite{LIGO:2021ppb}.  Given the new unrestricted bank, we used the configuration settings appropriate to the BBH search from \cite{LIGOScientific:2021djp}, with slight refinements.  We used a slightly different $\chi^{2}$ tuning which makes sure that no template uses less than 11 $\chi^{2}$ bins \cite{nitz20234}, and a low-frequency cutoff $f_\mathrm{low}$ of 10\,Hz is applied for the matched filter SNR evaluation across all detectors.  
The observation of repeated (lower amplitude) ``echo'' glitches, a few seconds away from loud auto-gated glitches which are windowed to zero~\cite{usman2016pycbc}, prompted the implementation of an additional veto \cite{chandra2021optimized}. This veto is applied to all triggers between $1.5$ seconds before and $2.5$ seconds after the central times of the gates for H1 and L1. For V1, triggers within $0.5$ seconds before and 1 second after the central gate times are excluded. These time intervals are selected based on visual inspection of the distribution of single-detector triggers around gated times. 
Following these settings, the PyCBC search is run with the \ac{KDE} ranking statistic.

\subsection{Search sensitivity comparisons}
\label{sec:search_sensitivity}
\begin{figure*}[tbp]
    \centering
    \begin{minipage}{\textwidth}
        \centering
        \includegraphics[scale=0.12]{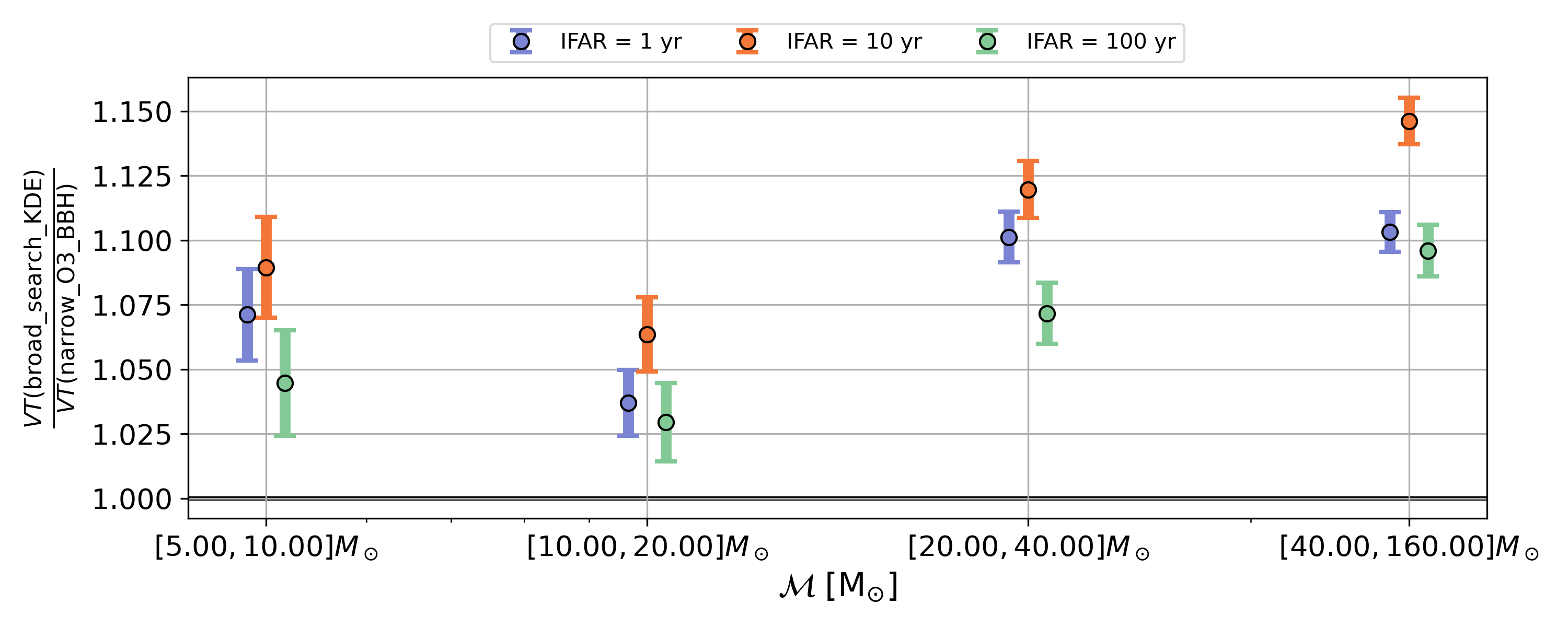}
    \end{minipage}
    \begin{minipage}{\textwidth}
        \centering
        \includegraphics[scale=0.178]{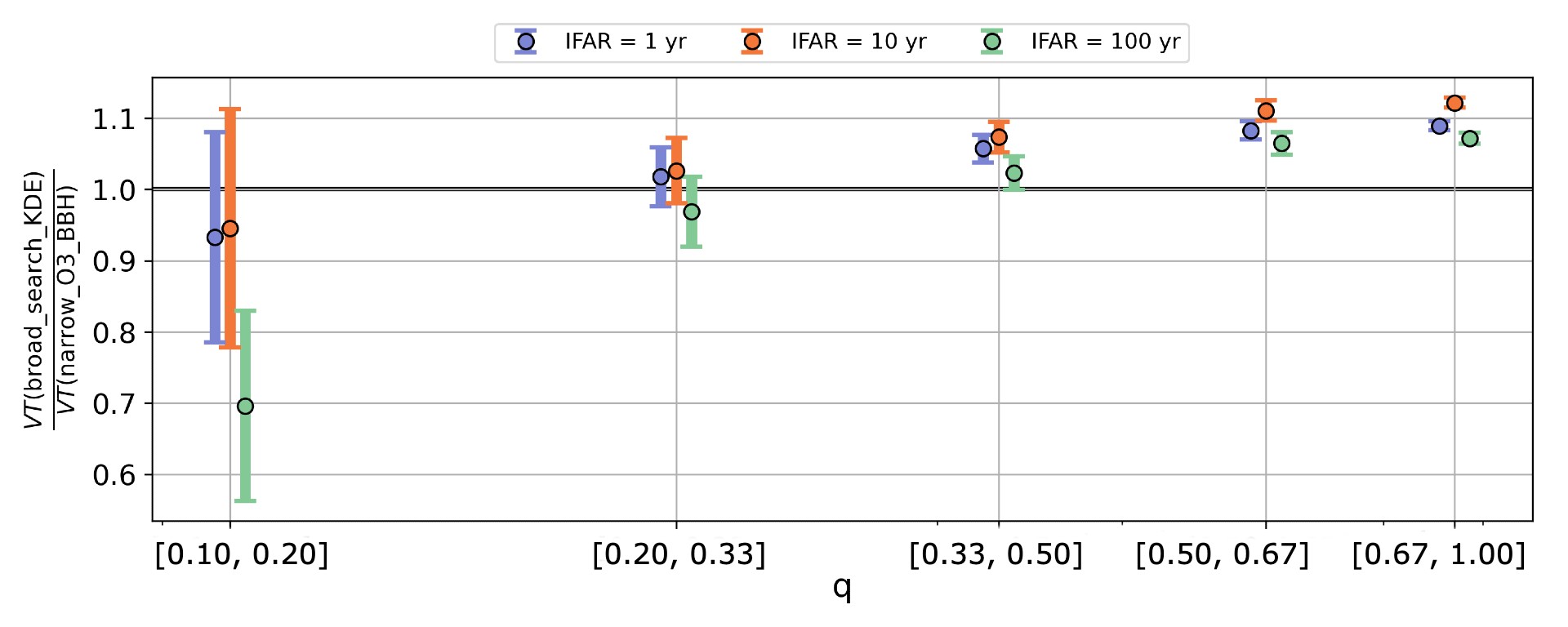}
    \end{minipage}
    \caption{Comparison of search sensitivity between two searches: \ac{KDE} (broad) and O3 \ac{BBH} (narrow), using the same set of injections. The top and bottom subplots show $\langle VT \rangle$ comparisons for injections divided into chirp mass bins and mass ratio bins respectively, at different \ac{IFAR} thresholds.}
    \label{vt}
\end{figure*}

To estimate the sensitive volume-time product $\langle VT \rangle$ of the searches, a large set of injections (simulated signals) are added to the data; we then count how many are recovered, i.e., detected by the search with an \ac{IFAR} above a specified threshold. 
The anticipated number of detections for a search is $\hat{N} = \langle VT \rangle R$, where $R$ denotes the astrophysical rate of mergers per unit volume and unit observing time. The injection component masses and spins are distributed within the range of 2 to 100\,$\msun$, and between -$0.998$ to $0.998$, respectively; their distribution, detailed in \cite{LIGOScientific:2021djp,ligo_scientific_collaboration_and_virgo_2023_7890437}, is primarily characterized by favoring more equal masses. % making unequal masses relatively uncommon in the injection set.  
The mass distribution follows separate power laws for each component mass: $p(m_1) \sim m_1^{-2.35}$ and $p(m_2 \mid m_1) \sim m_2$, while the redshift distribution follows $p(z) \sim dV_c / dz$, where $V_c(z)$ is the comoving volume out to redshift $z$.\footnote{This redshift distribution corresponds to a comoving rate density that \emph{increases} as $(1+z)^1$: this factor cancels in $p(z)$ against a $(1+z)^{-1}$ factor due to time dilation~\cite{LIGOScientific:2016kwr}.}
%\cite{fishbach2018does}. 
Injections are generated using the SEOBNRv4PHM waveform model~\cite{Ossokine:2020kjp}. Given the mass range of our bank, we consider only injections with $\mchirp \geq 5\,M_{\odot}$. %for sensitivity comparisons. 

Figure~\ref{vt} shows a comparison of injection sensitivities between the \ac{KDE}-based (broad) and O3 \ac{BBH} (narrow) searches over the entire O3 data set.  
These searches exhibit different characteristics, due to the nature of their respective template banks and the ranking statistic employed. 
In the case of O3 \ac{BBH}, the templates are narrowly concentrated around equal masses~\cite{Roy:2017oul} with mass ratios $q$ ranging from 1/3 to 1; in contrast, the \ac{KDE} search encompasses a broader parameter space with a wide range of mass ratios. 
%\td{These sentences repeat what you have already said in discussing the bank generation. Can you shorten this to not mention the component mass and spin ranges and just focus on mass ratio?}
%
%We see higher injection sensitivity in the KDE search:  the wider parameter coverage may increase sensitivity by enhancing the chances of detecting astrophysical phenomena across various parameter regimes. 
%
%\td{this sentence doesn't really say anything, it basically says we increase sensitivity by having a higher probability of detection, but that is the definition of sensitivity.}
The purpose of using a broader bank is to achieve a high match, and so a high recovered SNR, for signals with a wider range of parameters.
While the broader mass ratio range contributes to this sensitivity improvement, it might also result in a higher noise background, due to the larger number of templates. 
However, the \ac{KDE} statistic effectively penalizes templates in regions with a high density of noise events
allowing a focus on more promising areas. This enhances our ability to detect \ac{GW}s, even when injections don't precisely match any astrophysical distributions. Despite injections exhibiting a relatively smooth distribution, rather than following a detailed astrophysical population model, the \ac{KDE} statistic 
enhances sensitivity to these injections. 

Figure~\ref{vt} (bottom) compares injection sensitivities across mass ratio bins at various \ac{IFAR} thresholds. This comparison addresses a potential question regarding whether our higher sensitivity, observed over $\mchirp$ bins, can only be attributed to the bank covering a broader range of mass ratios. We observe lower sensitivity at low $q$ values; this may be expected, as our \ac{KDE} signal model disfavours unequal-mass systems.  The corresponding injections may still be detected in the O3 BBH search, with biased template parameters, while they can be detected in our search with low $q$ (or high spin) templates, resulting in a lower signal KDE and thus a lower ranking.  Conversely, in the range of mass ratios that is fully covered by the O3 BBH bank, the \ac{KDE} statistic still yields higher sensitivity; this must be due to the more detailed modelling of the template and signal densities by \acp{KDE}.

\subsection{Search Results}
\label{sec:search_results}

\begin{figure*}[tbp]
\centering
\begin{minipage}{0.465\textwidth} 
  \includegraphics[trim=5 1 130 1, clip,width=\textwidth]{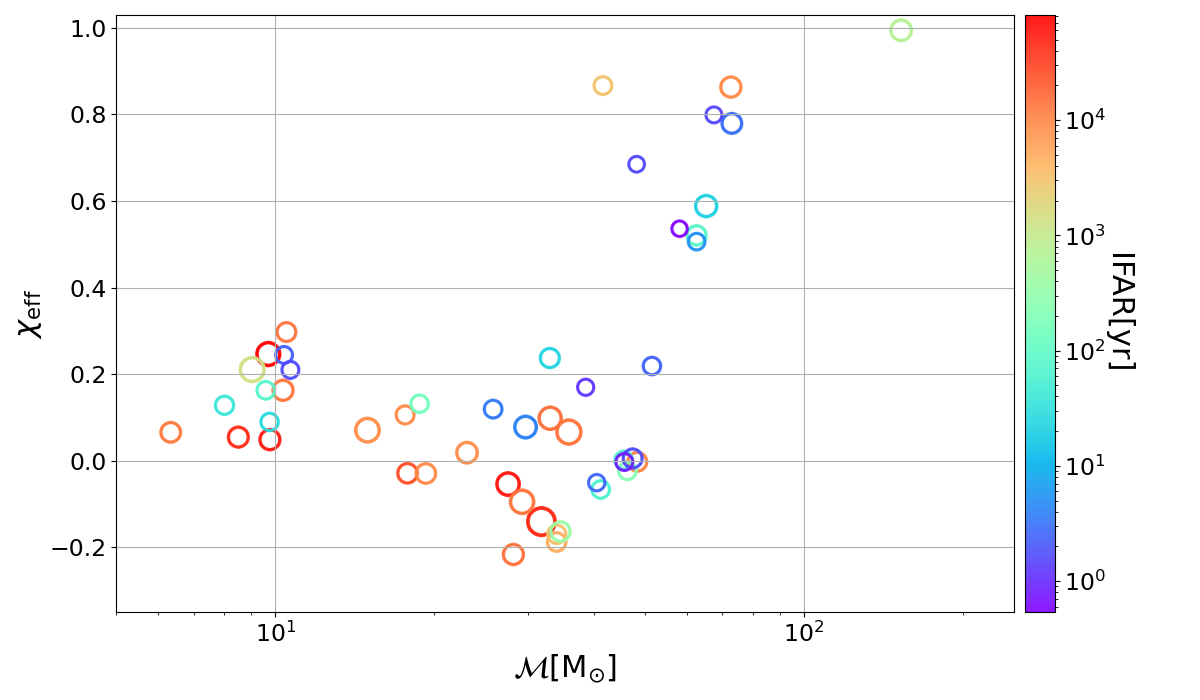}
\end{minipage}
\begin{minipage}{0.525\textwidth}
  \includegraphics[trim=5 1 45 1, clip,width=\textwidth]{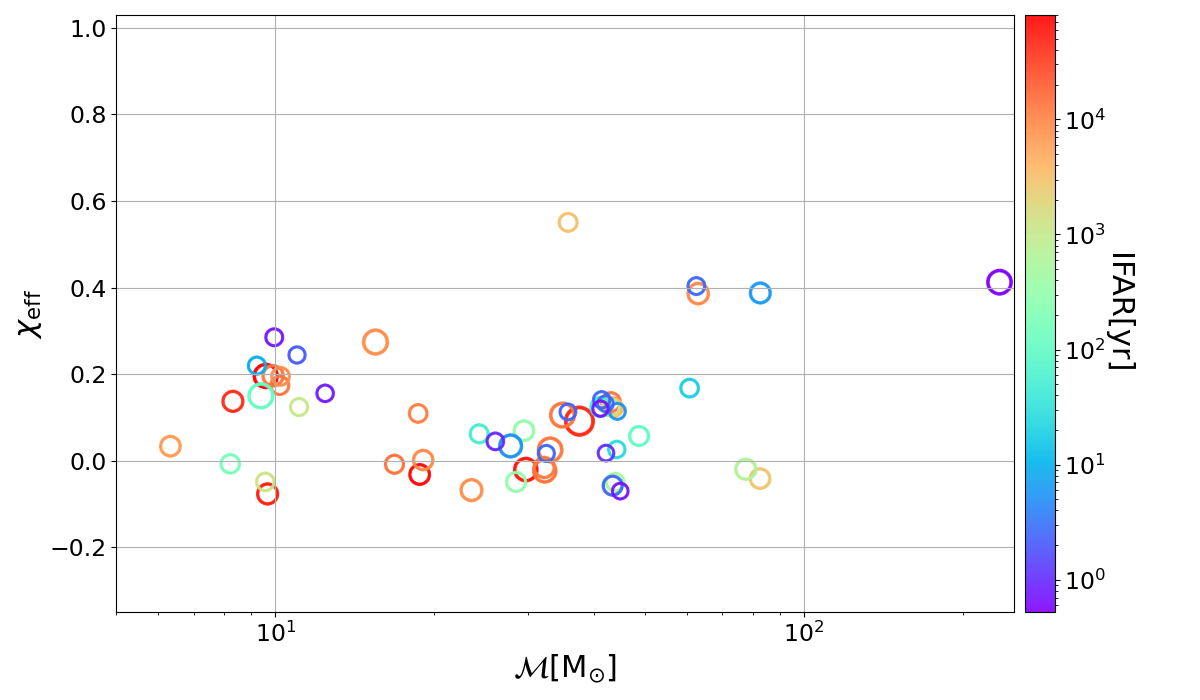}
\end{minipage}
\caption{Events detected with a threshold of \ac{IFAR} $> 0.5\,$yr by two different searches: O3 \ac{BBH} (left) and our unrestricted BBH search using the \ac{KDE} statistic (right). The color bar represents \ac{IFAR}, and the size of the symbols corresponds to the network \ac{SNR}.
}
\label{ifar}
\end{figure*}

We now discuss the set of events detected by our new \ac{KDE} search and compare it with other searches for \ac{BBH} mergers over the O3 data set. Initially, the PyCBC O3 \ac{BBH} search used in GWTC-3 identified a total of 49 events with an \ac{IFAR} exceeding 0.5 years, with a total probability of astrophysical origin ($p_{\rm astro}$) over these events of 48.33. Using the unrestricted bank and our new \ac{KDE} statistic, we see a number of detections increased by over $10\%$, with 57 events meeting the same threshold of \ac{IFAR} $> 0.5$ yr, and with a total astrophysical probability of 55.04; Fig.~\ref{ifar} illustrates the events detected by both searches. A total of 62 events have $p_{\rm astro}$ greater than 0.5 in our search. Additionally, the \ac{KDE} search outperforms the 4-OGC search~\cite{nitz20234} which reported 50 \ac{BBH} events with \ac{IFAR} $> 0.5$\,yr. Our event set is reported in Tables~\ref{golden} (high purity) and \ref{silver} (marginal), which also includes some candidates with \ac{IFAR} $< 0.5$\,yr but having a high probability of astrophysical origin. 

Since our signal model is trained on a sample of previous \ac{BBH} detections, many from O3, one might suspect the result for O3 data to be biased. Specifically, the statistic might be up-ranking the exact templates which correspond to known O3 events, rather than reflecting the underlying population distribution.  However, examining the signal \ac{KDE} in Fig.~\ref{densities}, it varies smoothly and slowly over the region where many detections are present; thus it is not ``over-fitting'' to individual O3 events, and we expect the ranking to be stable under small changes in the signal training set.  The fact that the search is able to detect several events that are \emph{not} in the training set, coupled with the increased sensitivity for an independent injection set, confirms a real sensitivity improvement.  

We find a conspicuous concentration of events, as well as higher sensitivity (see Fig.~\ref{vt}) near $\mchirp \approx 40\,\msun$, % this apparent correlation is 
consistent with the higher event count being a consequence of improved sensitivity.  Several events previously reported with \ac{IFAR} values below $0.5$ yr now surpass this threshold: 190527\_092055, 190725\_174728, 190916\_200658, 190926\_050336, 191127\_050227, 191224\_043228 and 200220\_124850.
However one \ac{BBH} event (200209\_085452) reported in \cite{KAGRA:2021duu} is not observed in our results with \ac{IFAR} $> 0.5\,$yr; we find a lower significance, possibly due to elevated $\chi^2$ values indicating a relatively poor match to available templates.  Events 190514\_065416, 191208\_080334, and 200301\_211019 from GWTC-3 also remain with \ac{IFAR} below $0.5\,$yr. 

Now considering events that were not seen in previous PyCBC BBH searches, 190711\_030756 and 200214\_223306 were first reported in \cite{olsen2022new} and \cite{mehta2023new}, respectively.  The IMBH marginal candidate event 191225\_215715, first reported in \cite{abbott2022search}, is also identified in our search, although with a low ($<0.5$) probability of astrophysical signal origin.  It stands out from others as its masses are significantly above the BBH-like events we use for training the signal \ac{KDE}. 
The signal \ac{KDE} is still nonzero outside the mass range of previously detected \acp{BBH}, but the exact values that determine the ranking of this event are strongly dependent on the bandwidth and other configuration choices. Since there is no real constraint on the signal model beyond the trained region, the significance estimate for 191225\_215715 may be affected by a---currently unknown---systematic bias. 
Adjustments to the \ac{KDE} statistic to allow for possible detections outside the previously observed population range are discussed further in Section~\ref{sec:discussion}. 

A nonzero $\chi_\mathrm{eff}$ indicates the definite presence of one or more component spins in the system, either aligned with the orbital angular momentum if positive, or anti-aligned if negative. Most detected binaries up to O3 are consistent with $\chi_\mathrm{eff} \approx 0$ \cite{LIGOScientific:2021usb, LIGOScientific:2021djp}. 
However, several events are identified in the O3 \ac{BBH} search by templates with high (positive) $\chi_\mathrm{eff}$, which appears inconsistent with \ac{PE} analyses, or at least indicates a significant bias in the search template parameters. 
Using the \ac{KDE} ranking statistic which incorporates an estimate of the signal spin distribution, we find that events now tend to cluster at lower $\chi_\mathrm{eff}$ values, narrowing the overall range of values.  Thus, in addition to optimizing search sensitivity, the KDE-based ranking may also improve the accuracy of the search template masses and spins in comparison to the credible regions found by \ac{PE} analysis. 

\begin{figure}[tbp]
%\centering
\hspace{-0.45cm}
\includegraphics[width=0.9\columnwidth]{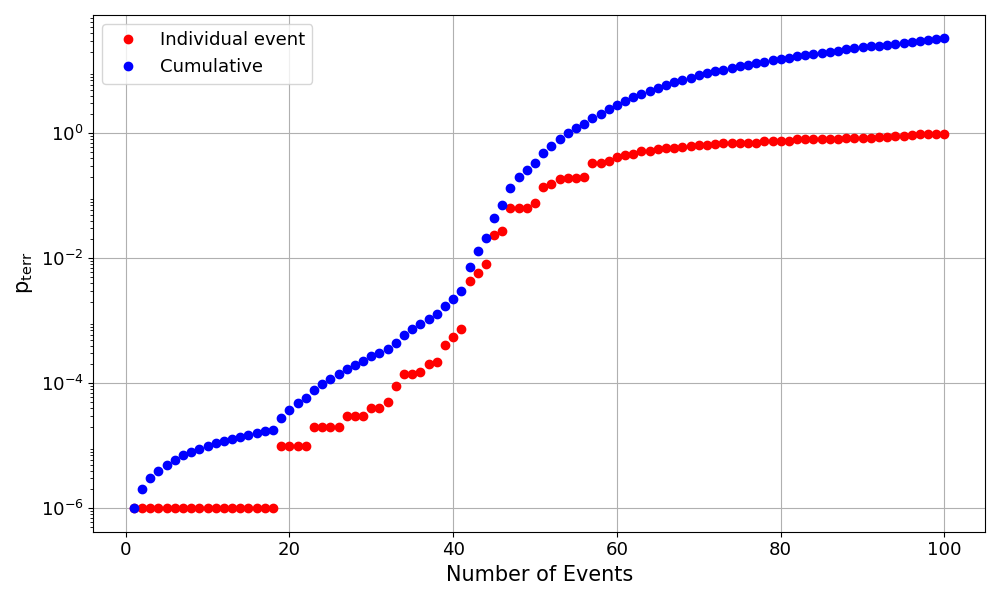}
\caption{Probability of terrestrial (noise) origin for the top 100 \ac{GW} event candidates. The red curve indicates the individual $p_\mathrm{terr}$ values, sorted in increasing order; lower values are considered more significant. The blue curve shows the cumulative terrestrial probability up to a given event, indicating the overall purity of the set of events.}
\label{pterr}
\end{figure}

\paragraph{Probability of astrophysical origin and catalog purity} 
Given the list of candidate \ac{GW} events from the search, together with the estimated distributions of noise events and astrophysical signal events over the search ranking statistic, we apply Bayesian inference on a mixture model, with the total number of signals present in the search results (including possible high-\ac{FAR} events) as an unknown parameter~\cite{Farr:2013yna,LIGOScientific:2016ebi,T1700029}.  Using this signal rate, and marginalizing over its uncertainty, we calculate the probability of astrophysical origin $p_\mathrm{astro}$ and of terrestrial (noise) origin $p_\mathrm{terr} = 1 - p_\mathrm{astro}$ for each candidate.  Highly significant events with low \ac{FAR} thus have $p_\mathrm{terr} \ll 1$. We calculated the cumulative $p_\mathrm{terr}$, i.e., the cumulative sum of $p_\mathrm{terr}$ values, sorted in \emph{increasing} order, over all our detected events.  This cumulative sum provides an overall idea of the likely noise contamination in a given set of events when looked at together, in contrast to the individual $p_\mathrm{terr}$ values. 
These cumulative $p_\mathrm{terr}$ values can be used to define event sets with more or less high purity, defined as the probable fraction of events that are astrophysical signals.

Thus, given the values shown in Fig.~\ref{pterr}, we defined two event sets based on significance and overall purity. In both tables, events are selected based on the criteria of \ac{IFAR} exceeding 0.5 years or a $p_\mathrm{astro}$ value greater than 0.5. 
Table~\ref{golden} then contains events with the smallest $p_\mathrm{terr}$ values, up to a cumulative $p_\mathrm{terr} < 1$, which we may call a `Gold' event set. Table~\ref{silver} then includes the remaining events for which the cumulative $p_\mathrm{terr} \geq 1$, known as the `Silver' event set. 
This classification % and presentation 
provides valuable insights into the observations over 
%\ac{GW} events from 
O3, helping us define event sets of well defined purity which may be used for further investigation into astrophysical or cosmological properties.

%\newpage

\onecolumngrid

%\newgeometry{left=2cm, right=2cm}
\begin{table}[h!]
\centering
\setlength{\arrayrulewidth}{0.5pt} % Set column rule width to zero to remove gaps
\setlength{\tabcolsep}{0.5pt} % Set column separation to zero
%\arrayrulecolor{cyan}               % Set the color of the column separators
\begin{tabular}{c c c c c S c c S} 
 \hline
\rowcolor{lightgray} Event & GPS & $m_{1}\,[\msun]$ & $m_{2}\,[\msun]$ & $\mchirp\,[\msun]$ & $\chi_\mathrm{eff}$ & $p_{\mathrm{astro}}$ & IFAR\,[yr] & \ac{SNR} \\ 
\hline

190408\_181802 & 1238782700.29  & 30.4     & 23.9     & 23.4     & -0.05    & 1.000    &    8.3e+03 & 14.09      \\
190412\_053044 & 1239082262.17  & 34.9     & 9.60     & 15.3     & 0.27     & 1.000    &    8.3e+03 & 18.26      \\
190413\_052954 & 1239168612.50  & 51.4     & 46.6     & 42.6     & 0.12     & 0.994    &        2.5 & 8.66       \\
190413\_134308 & 1239198206.74  & 89.5     & 61.5     & 64.4     & 0.41     & 0.996    &        2.7 & 9.16       \\
190421\_213856 & 1239917954.26  & 67.7     & 37.7     & 43.6     & -0.05    & 1.000    &    4.8e+02 & 10.12      \\
\rowcolor{black!10} 190503\_185404 & 1240944862.30  & 51.4     & 46.6     & 42.6     & 0.12     & 1.000    &        53. & 11.83      \\
\rowcolor{black!10} 190512\_180714 & 1241719652.42  & 24.5     & 18.9     & 18.7     & -0.00    & 1.000    &    9.1e+03 & 12.17      \\
\rowcolor{black!10} 190513\_205428 & 1241816086.75  & 34.8     & 31.8     & 28.9     & -0.04    & 1.000    &    3.2e+02 & 11.72      \\
\rowcolor{black!10} 190517\_055101 & 1242107479.83  & 48.0     & 36.4     & 36.3     & 0.57     & 1.000    &    3.0e+03 & 10.36      \\
\rowcolor{black!10} 190519\_153544 & 1242315362.40  & 89.5     & 61.5     & 64.4     & 0.41     & 1.000    &    9.1e+03 & 13.40      \\
190521\_030229 & 1242442967.46  & 101.     & 82.6     & 79.4     & -0.03    & 1.000    &    6.3e+02 & 12.94      \\
190521\_074359 & 1242459857.47  & 57.4     & 33.7     & 38.0     & 0.10     & 1.000    &    4.3e+04 & 24.36      \\
\textbf{190527\_092055} & 1242984073.79  & 42.3     & 34.2     & 33.1     & 0.02     & 0.937    &        2.6 & 8.37       \\
190602\_175927 & 1243533585.10  & 101.     & 82.6     & 79.4     & -0.03    & 1.000    &    2.7e+03 & 12.33      \\
190630\_185205 & 1245955943.18  & 48.4     & 21.4     & 27.5     & 0.02     & 1.000    &        5.4 & 15.44      \\
\rowcolor{black!10} 190701\_203306 & 1246048404.58  & 67.7     & 37.7     & 43.6     & -0.05    & 1.000    &        2.6 & 11.53      \\
\rowcolor{black!10} 190706\_222641 & 1246487219.33  & 115.     & 82.0     & 84.2     & 0.37     & 1.000    &        6.3 & 12.74      \\
\rowcolor{black!10} 190707\_093326 & 1246527224.17  & 12.5     & 10.3     & 9.89     & -0.06    & 1.000    &    5.3e+04 & 12.93      \\
\rowcolor{black!10} 190719\_215514 & 1247608532.93  & 45.5     & 37.0     & 35.7     & 0.12     & 0.937    &        2.3 & 8.09       \\
\rowcolor{black!10} 190720\_000836 & 1247616534.71  & 14.6     & 9.95     & 10.4     & 0.19     & 1.000    &    1.3e+04 & 10.64      \\
\textbf{190725\_174728} & 1248112066.47  & 21.5     & 5.43     & 9.00     & 0.21     & 0.977    &        9.1 & 9.42       \\
190727\_060333 & 1248242631.99  & 51.4     & 46.6     & 42.6     & 0.12     & 1.000    &    1.3e+04 & 11.52      \\
190728\_064510 & 1248331528.53  & 14.4     & 9.60     & 10.2     & 0.20     & 1.000    &    1.3e+04 & 13.24      \\
190731\_140936 & 1248617394.64  & 51.4     & 46.6     & 42.6     & 0.12     & 0.937    &        2.8 & 7.86       \\
190803\_022701 & 1248834439.88  & 64.0     & 38.6     & 43.0     & 0.02     & 0.992    &        27. & 8.69       \\
\rowcolor{black!10} 190828\_063405 & 1251009263.76  & 43.7     & 30.7     & 31.8     & -0.02    & 1.000    &    1.4e+04 & 15.90      \\
\rowcolor{black!10} 190828\_065509 & 1251010527.89  & 24.0     & 16.7     & 17.4     & -0.02    & 1.000    &    1.4e+04 & 10.53      \\
\rowcolor{black!10} 190915\_235702 & 1252627040.70  & 43.7     & 30.7     & 31.8     & -0.02    & 1.000    &    1.2e+04 & 13.07      \\
\rowcolor{black!10} 190924\_021846 & 1253326744.84  & 10.0     & 5.58     & 6.46     & 0.05     & 1.000    &    6.6e+03 & 12.38      \\
\rowcolor{black!10} 190925\_232845 & 1253489343.13  & 22.4     & 20.7     & 18.8     & 0.12     & 1.000    &    1.1e+04 & 10.06      \\
190930\_133541 & 1253885759.24  & 15.7     & 8.30     & 9.85     & 0.18     & 1.000    &    8.2e+03 & 10.11      \\
191105\_143521 & 1256999739.93  & 12.7     & 9.59     & 9.60     & -0.03    & 1.000    &    1.2e+03 & 9.94       \\
191109\_010717 & 1257296855.23  & 34.0     & 32.1     & 28.8     & 0.05     & 0.999    &    3.3e+02 & 12.41      \\
191126\_115259 & 1258804397.63  & 18.5     & 9.11     & 11.2     & 0.24     & 0.924    &        2.0 & 8.54       \\
\textbf{191127\_050227} & 1258866165.56  & 51.4     & 46.6     & 42.6     & 0.12     & 0.973    &        5.2 & 8.25       \\
\rowcolor{black!10} 191129\_134029 & 1259070047.20  & 13.6     & 7.22     & 8.53     & 0.13     & 1.000    &    4.1e+04 & 12.88      \\
\rowcolor{black!10} 191204\_110529 & 1259492747.54  & 41.5     & 23.0     & 26.7     & 0.06     & 0.816    &       0.97 & 8.97       \\
\rowcolor{black!10} 191204\_171526 & 1259514944.09  & 14.4     & 8.75     & 9.71     & 0.19     & 1.000    &    8.1e+04 & 17.20      \\
\rowcolor{black!10} 191215\_223052 & 1260484270.34  & 35.0     & 23.9     & 25.1     & 0.08     & 1.000    &        53. & 10.42      \\
\rowcolor{black!10} 191216\_213338 & 1260567236.48  & 10.4     & 10.4     & 9.04     & 0.13     & 1.000    &    1.2e+02 & 18.33      \\
191222\_033537 & 1261020955.12  & 92.7     & 38.0     & 50.6     & 0.05     & 1.000    &    1.0e+02 & 11.55      \\
\textbf{191224\_043228} & 1261197166.15  & 20.8     & 9.71     & 12.2     & 0.15     & 0.860    &       0.83 & 8.84       \\
191230\_180458 & 1261764316.41  & 90.5     & 57.0     & 62.2     & 0.16     & 0.999    &        19. & 10.09      \\
200128\_022011 & 1264213229.91  & 51.4     & 46.6     & 42.6     & 0.12     & 1.000    &    9.9e+03 & 9.23       \\
200129\_065458 & 1264316116.42  & 34.8     & 31.8     & 28.9     & -0.04    & 1.000    &    5.9e+04 & 16.08      \\
\rowcolor{black!10} 200202\_154313 & 1264693411.56  & 10.2     & 8.51     & 8.12     & -0.01    & 1.000    &    1.7e+02 & 11.05      \\
\rowcolor{black!10} 200208\_130117 & 1265202095.95  & 51.4     & 46.6     & 42.6     & 0.12     & 1.000    &    1.2e+03 & 9.75       \\
\rowcolor{black!10} 200219\_094415 & 1266140673.20  & 51.4     & 46.6     & 42.6     & 0.12     & 1.000    &    3.3e+03 & 9.54       \\
\rowcolor{black!10} \textbf{200220\_124850} & 1266238148.16  & 64.0     & 38.6     & 43.0     & 0.02     & 0.846    &        1.1 & 7.96       \\
\rowcolor{black!10} 200224\_222234 & 1266618172.40  & 45.5     & 37.0     & 35.7     & 0.12     & 1.000    &    1.3e+04 & 18.09      \\
200225\_060421 & 1266645879.40  & 21.1     & 20.6     & 18.2     & -0.05    & 1.000    &    7.2e+04 & 12.51      \\
200311\_115853 & 1267963151.39  & 42.3     & 34.2     & 33.1     & 0.02     & 1.000    &    1.3e+04 & 17.74      \\
200316\_215756 & 1268431094.16  & 17.3     & 8.81     & 10.6     & 0.13     & 1.000    &    1.1e+03 & 9.50       \\  \hline
\end{tabular}
\caption{\ac{GW} candidate events with either \ac{IFAR} or $p_\mathrm{astro}$ value greater than 0.5 yr from our \ac{KDE}-based broad \ac{BBH} search of O3 data, sorted by observation time.  The table presents a low contamination ``gold sample'' selected such that the cumulative $p_\mathrm{terr}$ is below $1$ (see main text). Events not found with \ac{IFAR} $> 0.5$ yr in previous PyCBC-based searches \cite{LIGOScientific:2021usb,LIGOScientific:2021djp, Nitz:2021uxj, nitz20234} are marked in bold. Template component and chirp masses are provided in detector frame.}
\label{golden}
\end{table}
%\restoregeometry

%\newgeometry{left=2cm, right=2cm}
\begin{table}[tbh]
\centering
\setlength{\arrayrulewidth}{0.5pt}     % Set column rule width to zero to remove gaps
\setlength{\tabcolsep}{0.5pt}          % Set column separation to zero

\arrayrulecolor{lightgray}
\begin{tabular}{c c c c c S c c S} 
\hline
\rowcolor{lightgray} Event & GPS & $m_{1}\,[\msun]$ & $m_{2}\,[\msun]$ & $\mchirp\,[\msun]$ & $\chi_\mathrm{eff}$ & $p_{\mathrm{astro}}$ & IFAR\,[yr] & \ac{SNR} \\ 
\hline

\textbf{190514\_065416} & 1241852074.85  & 67.7     & 37.7     & 43.6     & -0.05    & 0.583    &       0.21 & 8.07       \\
190711\_030756 & 1246849694.65  & 67.7     & 37.7     & 43.6     & -0.05    & 0.558    &       0.25 & 8.91       \\
\textbf{190916\_200658} & 1252699636.90  & 67.7     & 37.7     & 43.6     & -0.05    & 0.804    &       0.69 & 7.95       \\
\textbf{190926\_050336} & 1253509434.08  & 51.4     & 46.6     & 42.6     & 0.12     & 0.804    &       0.70 & 7.62       \\
190929\_012149 & 1253755327.51  & 109.     & 50.7     & 63.9     & 0.09     & 0.669    &       0.39 & 9.20       \\
\rowcolor{black!10} 191103\_012549 & 1256779567.53  & 12.8     & 10.6     & 10.1     & 0.30     & 0.803    &       0.76 & 9.25       \\
\rowcolor{black!10} \textbf{191208\_080334} & 1259827432.85  & 45.5     & 37.0     & 35.7     & 0.12     & 0.533    &       0.14 & 7.59       \\
\rowcolor{black!10} 191225\_215715 & 1261346253.87  & 337.     & 213.     & 232.     & 0.39     & 0.440    &       0.52 & 17.66      \\
\rowcolor{black!10} 200214\_223306 & 1265754805.00  & 90.5     & 57.0     & 62.2     & 0.16     & 0.647    &       0.32 & 7.80       \\
\rowcolor{black!10} \textbf{200301\_211019} & 1267132237.66  & 29.4     & 18.1     & 19.9     & -0.17    & 0.666    &       0.38 & 8.29       \\  
\hline
\end{tabular}
\caption{``Silver'' sample of marginal \ac{BBH} candidate events from O3 data, selected as in Table~\ref{golden} but with cumulative $p_\mathrm{terr} \geq 1$.}
\label{silver}
\end{table}

%\restoregeometry

\twocolumngrid

\section{Discussion}
\label{sec:discussion}

In this paper, we demonstrated a ranking statistic for gravitational-wave searches targeting compact binary mergers, which builds on previous work by using a model of the signal distribution over binary masses and spins based on previously detected significant events.  
The primary objective of this work is to combine the previous ``broad'' and ``BBH focused'' PyCBC searches employed for O3 data into one search only, which aims to increase sensitivity to the astrophysical population of signals, allowing for both the known high \ac{BBH} signal density and the detection of new types of signals.  This objective is achieved by incorporating \ac{KDE}-based models of the signal and template distributions into the search ranking statistic.
Conducting the PyCBC search using our developed model on the O3 data, we found around $10\%$ more events and also higher sensitivity compared to the O3 \ac{BBH} search. 

The basic approach of Eq.~\eqref{eq:R_and_R0}, i.e.\ introducing a signal density model in the ranking, may be extended or modified in various ways.  
One possible extension addresses the issue that previous detections, being a finite sample of events, do not give a complete picture of the full signal distribution.  Specifically, our choice of statistic will strongly down-rank any signals with parameters significantly outside the range of masses and spins covered by O1-O3 \ac{BBH} detections.  A possible remedy is to limit how far any template may be down-ranked, which we can achieve by adding an extra term to the signal density in order to represent possible events with parameters outside the previously seen distribution.  Formally, the signal model is then a mixture of the KDE based on some set of known events with a ``broad'', (near-)constant density covering the entire search space (or bank).  We may write
\begin{equation}
    \hat{h}(x) = (1 - a) \hat{f}(x) + a C, 
\end{equation}
where $a$ is the mixture fraction of the ``broad'' component and $C$ is a normalization constant: $C = 1/v$, where $v$ is the volume of the parameter space covered by the template bank in the coordinates used for the \ac{KDE}.  The parameter $a$ may be adjusted to give more or less weight to the detection of any ``exceptional'' events outside the known population. 

One may also take a signal density based on a more or less realistic formal or astrophysical model, rather than empirically on previous detections.  The signal model may be an arbitrary function, but in some cases can be obtained by applying a KDE: the estimate will use as input a set of \emph{simulated} detectable events (i.e.\ those above a fixed \ac{SNR} threshold) whose distribution follows the desired model.  The simulated events may derive from an astrophysical calculation, for instance stellar population synthesis (e.g.~\cite{Dominik:2014yma}) or modelling gravitationally lensed counterparts~\cite{li2023tesla}, or from other sampling procedures.  The KDE method is particularly applicable for complicated source distributions which cannot be written in closed form, or where only a relatively small number of sample events may be available. 

The application of our optimized adaptive KDE to relatively large data sets (order $10^5$ points or more) can be computationally demanding.  Determining the best choice of bandwidth and adaptive parameter by grid search requires some tens of KDE evaluations, becoming impractical if the time for a single KDE on a CPU core exceeds about an hour.  While parallelization could be of benefit, some change in method is required for the template banks used in current searches over a full parameter space of stellar mass binaries, with up to $\mathcal{O}(10^6)$ points. 
One possible strategy to address computational cost for large banks is downsampling the input points, with the probability of rejection being a known function of template parameters (for instance, a power of chirp mass): after the KDE is computed from a reduced set of points, the density may then be corrected for the rejection step.  Alternatively, the bank may be split into a number of sub-banks with a separate KDE evaluated over each one; while this approach is easily parallelizable, care is required in combining the sub-bank KDEs in order to avoid edge effects or other artefacts. 

We may also consider applying a KDE-based statistic in a context beyond the state of the art of \ac{CBC} searches over binary masses and orbit-aligned spins, with templates representing only the dominant emission multipole.  When the search is extended to templates including subdominant multipole emission (``higher modes''), it is necessary to construct the bank over a space with additional dimensions representing the binary orientation, in particular the orbital inclination angle~\cite{Harry:2017weg,Chandra:2022ixv}.  Similarly, a search including the orbital precession effects of non-aligned spins requires a larger number of degrees of freedom in bank construction, for instance the spin magnitude and opening angle used in~\cite{Pan:2003qt}, or the amplitudes and phases of harmonic waveform components in~\cite{McIsaac:2023ijd}.  In both the higher modes and precession cases, the distribution of templates and of signals over the higher-dimensional space may show nontrivial structure that is difficult to predict or model analytically; the adaptive KDE then offers a possible route to optimize the sensitivity of these more complex searches.

\section*{Acknowledgments}
We are grateful to Florian Aubin for carefully reading an earlier version of this paper, and to the PyCBC search development team for discussions during this project, and for making details of previous O3 searches available.  This work has received financial support from Xunta de Galicia (CIGUS Network of research centers), by European Union ERDF and by the ``Mar{\'i}a de Maeztu'' Units of Excellence program CEX2020-001035-M and the Spanish Research State Agency.  TD and PK are supported by research grant PID2020-118635GB-I00 from the Spanish Ministerio de Ciencia e Innovaci{\'o}n.  

This research project was made possible through the access granted by the Galician Supercomputing Center (CESGA) to its supercomputing infrastructure.  The authors are also grateful for computational resources provided by the LIGO Laboratory and supported by National Science Foundation Grants PHY-0757058 and PHY-0823459.  This material is based upon work supported by NSF’s LIGO Laboratory which is a major facility fully funded by the National Science Foundation.

\bibliographystyle{ieeetr}
\bibliography{references}

\end{document}